\documentclass[aps,prd,twocolumn,showpacs,floatfix,preprintnumbers,amsmath,amssymb, notitlepage, superscriptaddress,nofootinbib]{revtex4-1}

\input epsf
\usepackage{graphicx}
\usepackage{color}
\usepackage{mathtools}
\usepackage{mathbbol}

\newcommand{\beq}{\begin{equation}}
\newcommand{\eeq}{\end{equation}}
\newcommand{\barr}{\begin{eqnarray}}
\newcommand{\earr}{\end{eqnarray}}

\newcommand{\rme}{\textrm{e}}

\newcommand{\bs}{\boldsymbol}

\newcommand{\apjl}{Astrophys. J. Lett.}
\newcommand{\aap}{Astron. Astrophys.}
\newcommand{\apjs}{Astrophys. J. Suppl. Ser.}

\newcommand{\mnras}{Mon. Not. R. Astron. Soc.}

\newcommand{\physrep}{Phys. Rep.}
\newcommand{\araa}{Annual Rev. Astron. Astrophys.}

\newcommand{\pasj}{Publ. Astron. Soc. Japan}

\usepackage{color}
\newcommand{\yah}[1]{{#1}}
\newcommand{\caca}[1]{{#1}}

\newcommand{\lsim}{\mathrel{\hbox{\rlap{\lower.55ex\hbox{$\sim$}} \kern-.3em \raise.4ex \hbox{$<$}}}}
\newcommand{\gsim}{\mathrel{\hbox{\rlap{\lower.55ex\hbox{$\sim$}} \kern-.3em \raise.4ex \hbox{$>$}}}}

\begin{document}
\title{The merger rate of primordial-black-hole binaries}
\author{Yacine Ali-Ha\"imoud}
\affiliation{Center for Cosmology and Particle Physics, Department of Physics, New York University, New York, NY 10003, USA}
\author{Ely D. Kovetz}
\affiliation{Department of Physics and Astronomy, Johns Hopkins University, Baltimore, MD 21218, USA}
\author{Marc Kamionkowski}
\affiliation{Department of Physics and Astronomy, Johns Hopkins University, Baltimore, MD 21218, USA}
\date{\today}

\begin{abstract}

Primordial black holes (PBHs) have long been a candidate for the elusive dark matter (DM), and remain poorly constrained in the $\sim 20-100~ M_{\odot}$ mass range. PBH binaries were recently suggested as the possible source of LIGO's first detections. In this paper, we thoroughly revisit existing estimates of the merger rate of PBH binaries. We compute the probability distribution of orbital parameters for PBH binaries formed in the early Universe, accounting for tidal torquing by all other PBHs, as well as standard large-scale adiabatic perturbations. We then check whether the orbital parameters of PBH binaries formed in the early Universe can be significantly affected between formation and merger. Our analytic estimates indicate that the tidal field of halos and interactions with other PBHs, as well as dynamical friction by unbound standard DM particles, do not do significant work on nor torque PBH binaries. We estimate the torque due to baryon accretion to be much weaker than previous calculations, albeit possibly large enough to significantly affect the eccentricity of typical PBH binaries. We also revisit the PBH-binary merger rate resulting from gravitational capture in present-day halos, accounting for Poisson fluctuations. If binaries formed in the early Universe survive to the present time, as suggested by our analytic estimates, they dominate the total PBH merger rate. Moreover, this merger rate would be orders of magnitude larger than LIGO's current upper limits if PBHs make a significant fraction of the dark matter. As a consequence, LIGO would constrain $\sim 10-300 ~M_{\odot}$ PBHs to constitute no more than $\sim 1\%$ of the dark matter. To make this conclusion fully robust, though, numerical study of several complex astrophysical processes -- such as the formation of the first PBH halos and how they may affect PBH binaries, as well as the accretion of gas onto an extremely eccentric binary -- is needed.

\end{abstract}

\maketitle

\section{Introduction}

The nature of dark matter has eluded several generations of theoretical and experimental physicists. While there is no shortage of candidates, from ultra-light axion-like scalar fields \cite{Marsh_16}, to weakly interacting massive particles \cite{Jungman_96}, and massive compact objects \cite{Griest_91, Dolgov:2008wu}, dedicated dark-matter experiments remain stubbornly silent. As LIGO is slowly but surely ushering astronomy into the gravitational-wave era \cite{LIGO_16, GW151226, GW170104}, one of the oldest dark-matter candidates, primordial black holes (PBHs), has recently been brought back into the spotlight \cite{Bird_16, Sasaki_16, Cholis:2016kqi, Clesse:2016vqa, Raccanelli:2016cud, Kovetz_17, Blinnikov:2016bxu, Kashlinsky:2016sdv}. 

The notion of PBHs \yah{\cite{1967SvA....10..602Z} was first fleshed out by Ref. \cite{Hawking_71}}, who suggested that they may form out of large fluctuations in the early Universe, and may be of any mass above the Planck mass. The abundance of PBHs is bounded by a variety of observables, depending on their mass \cite{Carr_10, Carr_16}. In the $\sim 20-100~M_{\odot}$ mass range, PBHs are not constrained by microlensing \cite{Macho_01, Eros_07}, wide Galactic binaries \cite{Monroy_14}, nor the cosmic microwave background (CMB) \cite{Ricotti_08, YAH_17}. 
\yah{Refs.~\cite{Brandt_16, Koushiappas_17} have argued that dynamical heating of stellar systems in dwarf galaxies constrain compact objects in this mass range to make up no more than $\sim10\%$ of the dark matter, but Refs.~\cite{Brandt_16, Li:2016utv} note the possibility that intermediate-mass black holes in such systems would weaken this constraint. Others \cite{Gaggero:2016dpq,Inoue:2017csr} have set similar bounds in this mass range from the non-observation of radiation from accretion of interstellar gas onto such PBHs, but the radiative-feedback calculations upon which these conclusions rely are highly uncertain \cite{Agol:2001hb,YAH_17}; see also Refs.~\cite{Carr_17, Green_17, Poulin_17} for caveats on other bounds.}
As more data becomes available, it will be possible to constrain PBHs more tightly (e.g.~\cite{Garcia-Bellido_17, Kovetz:2016kpi, Kovetz_17, Munoz:2016tmg,Schutz:2016khr, Raidal_17}). In the meantime, it is important to investigate different avenues to probe PBHs with existing data.

Two decades ago, Nakamura \emph{et al.}\ \cite{Nakamura_97} (hereafter, NSTT) pointed out that PBHs would form binaries in the early Universe, which would then slowly shrink through gravitational wave radiation, and eventually coalesce. They estimated the merger rate per galaxy at the present time, and found it to be within the reach of the first gravitational-wave detectors. The LIGO Scientific  Collaboration searched for such mergers in LIGO's second science run \cite{LIGO_05}, but could only set upper bounds on the merger rate three to four orders of magnitude larger than NSTT's prediction if PBHs make all of the dark matter. A decade later, LIGO's sensitivity has increased spectacularly, pushing out the horizon distance to cosmological scales. It is straightforward to transpose NSTT's result to the $\sim 20-100~M_{\odot}$ range \cite{Sasaki_16}, and find that the predicted merger rate is glaringly larger than the latest estimate of the binary-black-hole merger rate from LIGO's three detections, $R \approx 10-200$ Gpc$^{-3}$ yr$^{-1}$ in the $\sim 5-100~ M_{\odot}$ mass range \cite{GW170104}. It is therefore clear that LIGO has the potential to significantly constrain PBH dark matter \cite{Raidal_17}, provided the rate estimated in NSTT is accurate. 

The main goal of the present work is to thoroughly check the merger rate of PBH binaries. We start, in Section \ref{sec:initial_cond}, by deriving the distribution of initial orbital parameters of PBH binaries forming in the early Universe, accounting for tidal torquing by \emph{all} other PBHs (as opposed to just the nearest neighbor as in NSTT), as well as by standard adiabatic perturbations. Some of the technical details of the calculation are deferred to the Appendix. Section \ref{sec:halos} is devoted to checking one of the key underlying assumptions of NSTT, namely that PBH binaries formed in the early Universe are subsequently only subject to gravitational radiation reaction. While Refs.~\cite{Ioka_98, Sasaki_16} have checked that PBH binaries do not get disrupted in present-day, Milky-Way-like halos, they did not consider the effect of the earliest non-linear structures. We fill that gap, estimating analytically the properties of the first halos and their effect on PBH binaries, which we find to be small. We also estimate the effect of baryon accretion, and find it to be at the verge of being relevant. In Section \ref{sec:bird}, we revisit the late-Universe PBH-binary formation mechanism proposed in Ref.~\cite{Bird_16}. While the merger rate obtained there is significantly lower than that due to early-Universe binaries, a key ingredient was omitted in their calculation. Indeed, when estimating the properties and mass function of halos, they did not account for the large Poisson fluctuations resulting from the \yah{granularity of PBH-dark-matter (although they did speculate that this granularity would increase the rates)}. Including these, we find that, while the merger rate per halo is much enhanced, the faster evaporation of denser halos truncates the mass function at a larger mass, leading to a final result that is comparable with that of Ref.~\cite{Bird_16}. In Section \ref{sec:limits} we derive potential upper limits on the PBH abundance from LIGO's existing upper bounds on the merger rate. 

\yah{As we discuss in the concluding section, our results suggest that LIGO severely constrains PBH dark matter in the $\sim10-300\,M_\odot$ mass range, eliminating them as the dominant component of the dark matter. To make this conclusion fully robust, though, requires careful numerical investigation of how the first PBH halos form and how PBH binaries survive in these halos, and investigation of the impact of baryon accretion onto a highly eccentric binary. Given these remaining open questions, continued pursuit of other observational probes of PBH dark matter in this mass range is still warranted.}

\section{PBH binary formation in the early Universe} \label{sec:initial_cond}

\subsection{Assumptions and notation}
Throughout this paper we use geometric units $G = c = 1$. Whenever relevant, we use cosmological parameters consistent with the latest \emph{Planck} measurements \cite{Planck_16}. We denote by $t_0 \approx 14$ Gyr the present time.

We denote by $s$ the scale factor normalized to unity at matter-radiation equality, and by $\rho_{\rm eq}$ the density of matter (or radiation) at equality. Neglecting dark energy and curvature, the Hubble rate is given by
\beq
H(s) = \left(\frac{8 \pi}{3} \rho_{\rm eq}\right)^{1/2} h(s), \ \ \ \ \ \ h(s) \equiv \sqrt{s^{-3} + s^{-4}}.
\eeq

We assume that PBHs make a fraction $f$ of the nonrelativistic matter (i.e.~dark matter and baryons as this simplifies expressions). We show our results in terms of the fraction $f_{\rm pbh} \approx f/0.85$ of PBHs in dark matter. We assume PBHs have a single mass $M =  m~ M_{\odot}$. All results carry over to a relatively narrow mass function, though more work would be required to generalize them to an extended mass function. 

We assume that the large-scale distribution of PBHs follows that of the dark matter, but that on sufficiently small scales, and at early enough times, they are effectively randomly distributed in space. In other words, we neglect the possibility of small-scale PBH clustering \cite{Chisholm_06}, which depends on the details of the PBH formation model. If PBHs do not make all of the dark matter, we denote by $\sigma_{\rm eq}^2$ the variance of density perturbations of the rest of dark matter \caca{on scales of order $\sim (10^{-3} - 10^3) ~M \sim (10^{-2} - 10^5)~ M_{\odot}$, at equality}. Extrapolating the measured amplitude and spectral index of primordial adiabatic perturbations \cite{Planck_16} to these small scales, one gets $\sigma_{\rm eq} \approx 0.005$, with a weak (logarithmic) dependence on mass. We adopt this value whenever required\footnote{\yah{It is worth pointing out that early-Universe scenarios for the formation of PBHs typically involve enhanced primordial power, in which case $\sigma_{\rm eq}$ may be significantly higher. We leave the examination of particular PBH formation models to future work.}}.

We denote by $\overline{x}$ the characteristic comoving separation (i.e.~proper separation at equality) between two PBHs, 
\beq
\overline{x} \equiv \left(\frac{3 M}{4\pi f \rho_{\rm eq}}\right)^{1/3}.
\eeq
Given a comoving separation $x$, we define the dimensionless variable $X$ as
\beq
X \equiv (x /\overline{x})^3.
\eeq
Provided a pair of PBHs are close enough, they decouple from the Hubble flow early on, and form a binary. We denote by $a$ the semi-major axis of a PBH binary, by $\ell$ its angular momentum per unit reduced mass, and by 
\beq
j \equiv \ell/\sqrt{2 M a} = \sqrt{1-e^2}
\eeq 
its dimensionless angular momentum, where $e \in [0, 1]$ is its eccentricity. We now estimate the probability distribution of these initial orbital parameters.

\subsection{Initial semi-major axis} \label{sec:sma}

Consider two point masses $M$, initially at comoving separation $x$, with vanishing peculiar velocity, in an otherwise homogenous expanding Universe. As long as the separation is much smaller than the Hubble scale, we may use a Newtonian approximation. If no perturber is present, the motion is one-dimensional. We denote by $r \in \mathbb{R}$ the proper separation projected along the axis of motion; it evolves according to
\barr
\ddot{r} - (\dot{H} + H^2 ) r + \frac{2 M}{r^2} \frac{r}{|r|} = 0, \label{eq:eom-t}
\earr
where overdots denote differentiation with respect to the proper time. We define $\chi \equiv r/x$ and rewrite Eq.~\eqref{eq:eom-t} in terms of the scale factor $s$: 
\barr
\chi'' + \frac{s h' + h}{s^2 h}  \left( s \chi' - \chi \right) + \frac{1}{\lambda} \frac1{( s h)^2} \frac{1}{\chi^2} \frac{\chi}{|\chi|} = 0, \label{eq:eom}
\earr
where primes denote differentiation with respect to $s$, and the dimensionless parameter $\lambda$ is
\beq
\lambda \equiv \frac{4 \pi \rho_{\rm eq} x^3}{3 M} = \frac{X}{f}. \label{eq:lambda}
\eeq
At $s \rightarrow 0$, the binary follows the Hubble flow $\chi(s) = s$, so the initial conditions are 
\barr
\chi(0) = 0,  \ \ \chi'(0) = 1.
\earr
We see that the solution is entirely characterized by $\lambda$.

\begin{figure}
\includegraphics[width = \columnwidth]{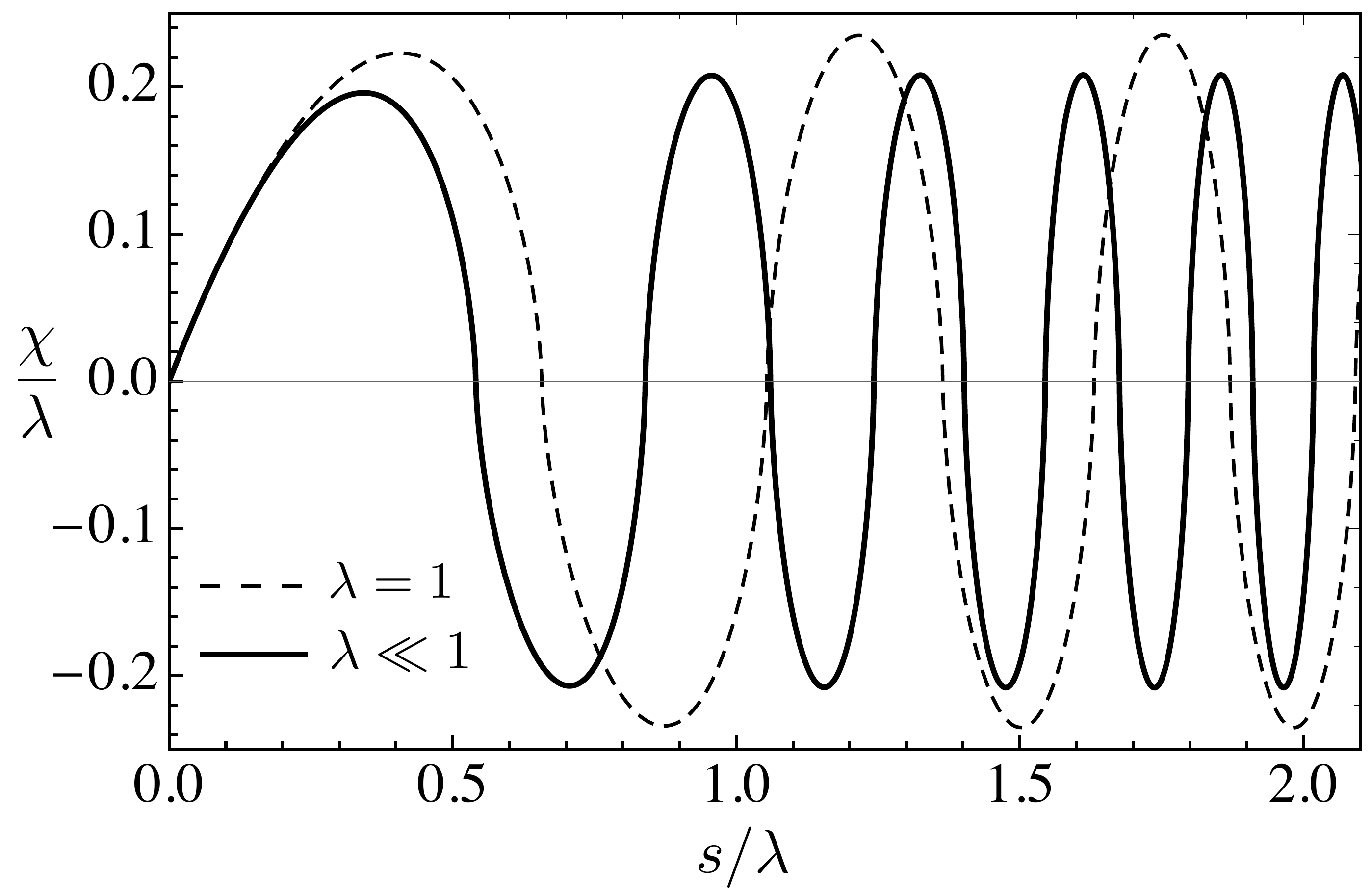}
\caption{Dimensionless separation $\chi = r/x$ of two point masses, rescaled by the parameter $\lambda = \frac1f (x/\overline{x})^3$, as a function of the rescaled scale factor $s/\lambda$, in the limit $\lambda \ll 1$ (solid) and for $\lambda = 1$ (dashed).} \label{fig:chi_lambda}
\end{figure}

In the limit $\lambda \ll 1$, the PBH pair effectively decouples from the expansion deep in the radiation-domination era, $s \ll 1$. In that limit, $h(s) \approx s^{-2}$, and the equation of motion is
\beq
\chi'' - \frac1{s^2} \left( s \chi' - \chi \right) +  \frac1{\lambda} \frac{s^2}{\chi^2} \frac{\chi}{|\chi|} = 0. \label{eq:chi-ODE-rad-dom}
\eeq
One can show that the solution to this equation is self-similar: 
\beq
\chi(s; \lambda) = \lambda ~\chi(s/\lambda; 1). \label{eq:self-similar}
\eeq
We compute this function numerically by solving Eq.~\eqref{eq:chi-ODE-rad-dom} and show it in Fig.~\ref{fig:chi_lambda}: we find that the binary effectively decouples from the Hubble flow at $s \approx \lambda/3$, and that the proper separation then oscillates with amplitude $|\chi| \approx 0.2~\lambda = 2 a/x$, where $a$ is the semi-major axis of the newly formed binary. We therefore find, for $\lambda \ll 1$,
\barr
a &\approx& 0.1~\lambda~x = \frac{0.1}{f} \frac{x^4}{\overline{x}^3} = 0.1 \left(\frac{3 M}{4 \pi \rho_{\rm eq}}\right)^{1/3} \left(X/f\right)^{4/3}. ~~~\label{eq:a_ov_x}
\earr
This agrees with the result of Ref.~\cite{Ioka_98} given that they define the mean separation without the factor of $(4 \pi/3)^{1/3}$. Solving the full equation \eqref{eq:eom}, we find that this result remains reasonably accurate even for $\lambda \sim 1$ (see Fig.~\ref{fig:chi_lambda}). In what follows we will see that for the PBH masses considered, the bulk of binaries merging at the present time have $\lambda \lesssim 1$, so we use Eq.~\eqref{eq:a_ov_x} throughout.

\subsection{Initial angular momentum} \label{sec:j_init}

We now account for the fact that the binary is immersed in a local tidal field $T_{ij} = -\partial_i \partial_j \phi$, which exerts a perturbative force per unit mass $\bs{F} = \mathbf{T} \cdot \bs{r}$, in matrix notation. This tidal field arises from the other PBHs, as well as from matter density perturbations, as pointed out in Ref.~\cite{Eroshenko_16} (see also \cite{Hayasaki_16}). Provided the initial comoving separation of the binary is small relative to the mean separation, this tidal field does not significantly affect the binary's energy (hence semi-major axis). However, it produces a torque $\dot{\bs{\ell}} = \bs{r} \times [\mathbf{T} \cdot \bs{r}]$, resulting in a non-vanishing angular momentum 
\beq
\bs{\ell} = \int dt ~\bs{r} \times [\mathbf{T} \cdot \bs{r}],
\eeq
and preventing a head-on collision. If the torque originates from other PBHs whose comoving separation is approximately constant (which is accurate provided their separation is much larger than $x$), then $\mathbf{T} \propto 1/s^3$. If the torque originates from linear matter density perturbations, then $T_{ij} \sim \rho_m \delta_m \propto s^{-3} \delta_m$. If the binary decouples deep in the radiation era, then $\delta_m \approx $ constant (neglecting the slow logarithmic growth). Therefore in either case, we have $\mathbf{T} \approx s^{-3} \mathbf{T}_{\rm eq}$. We hence get
\beq
\bs{\ell} = \left(\frac{3}{8 \pi \rho_{\rm eq}}\right)^{1/2} \int \frac{ds}{s h(s)} \frac{\chi^2(s; \lambda)}{s^3}  ~  \bs{x} \times   [\mathbf{T}_{\rm eq} \cdot \bs{x}].
\eeq
The integral only depends on $\lambda$. In the limit $\lambda \ll 1$, using the self-similarity relation \eqref{eq:self-similar}, it simplifies to
\beq
\int \frac{ds}{s h(s)} \frac{\chi^2(s; \lambda)}{s^3} = \lambda \int \frac{d \tilde{s}}{\tilde{s}^2} \chi^2(\tilde{s}; 1) \approx 0.3~\lambda,
\eeq
where we computed the last integral numerically. The reduced angular momentum $\bs{j} \equiv \bs{\ell}/\sqrt{2Ma}$ is therefore
\barr
\bs{j} &\approx& \frac{0.3}{\sqrt{0.1}} \lambda^{1/2}  \left(\frac{3}{16 \pi \rho_{\rm eq} M}\right)^{1/2} x^{3/2} \hat{x} \times \left[\mathbf{T}_{\rm eq} \cdot \hat{x}\right],\nonumber\\
&\approx& 0.5~ x^3~ \hat{x} \times \left[\frac{\mathbf{T}_{\rm eq} }{M}\cdot \hat{x}\right] \label{eq:j-general}
\earr
where we used Eqs.~\eqref{eq:lambda} and \eqref{eq:a_ov_x} to simplify the expression. 

\subsubsection{Torques by other PBHs}

Let us now specifically consider the tidal field generated by a point mass $M$ at comoving separation $y \gg x$:
\beq
\frac{T_{\rm eq}^{ij}}{M} = \frac{3 \hat{y}^i \hat{y}^j - \delta^{ij}}{y^3}.
\eeq 
This implies an angular momentum
\beq
\bs{j} \approx 1.5 \frac{x^3}{y^3} (\hat{x} \cdot \hat{y}) (\hat{x} \times \hat{y}),
\eeq
with magnitude $j \approx 0.8 (x/y)^3 \sin (2 \theta)$, where $\theta$ is the angle between $\hat{x}$ and $\hat{y}$, consistent with the results of Ref.~\cite{Ioka_98}.

The total reduced angular momentum resulting from \emph{all} other PBHs (at distance $y \gg x$) is hence given by
\beq
\bs{j} \approx 1.5 \sum_p \frac{x^3}{y_p^3} (\hat{x} \cdot \hat{y}_p) (\hat{x} \times \hat{y}_p). \label{eq:j-sum}
\eeq
We compute explicitly the probability distribution of $j$ in the Appendix, where we find, for a given $X$,
\barr
j \frac{dP}{dj}\Big{|}_X &=& \mathcal{P}(j/j_X), \ \ \ \ \mathcal{P}(\gamma) \equiv \frac{\gamma^2}{(1 + \gamma^2)^{3/2}}, \label{eq:dP_dj}\\
\textrm{with} \ \ \ \ \ \ j_X &\equiv& 0.5 X. \label{eq:j_X}
\earr
Note that this distribution extends to arbitrarily large $j$, while physical values are limited to $j \leq 1$. As long as $j_X \ll 1$, the contribution of unphysical values $j > 1$ is negligibly small. We emphasize that this probability distribution accounts for torques by \emph{all} PBHs. \caca{In contrast, Refs.~\cite{Nakamura_97, Ioka_98} only considered torques by the nearest neighbor, which leads to the correct approximate characteristic value of $j$, but does not allow to estimate its exact probability distribution.} 

\subsubsection{Torques by linear density perturbations} \label{sec:torques_dm}

As pointed out in Refs.~\cite{Hayasaki_16, Eroshenko_16}, if the PBH fraction is smaller than the characteristic large-scale matter density perturbation $\delta_m$, then tidal torques are dominated by large-scale linear perturbations, $T_{\rm eq}^{ij} = - \partial_i \partial_j \phi = - 4 \pi \rho_{\rm eq} \partial_i \partial_j \partial^{-2} \delta_m$. The resulting $\bs{j}$ is Gaussian-distributed in the plane perpendicular to $\hat{x}$, with variance given by [see Appendix \ref{sec:j2}]
\beq
\langle j^2 \rangle^{1/2} = \sqrt{\frac3{10}}  \frac{\sigma_{\rm eq}}f  X \approx 0.5 \frac{\sigma_{\rm eq}}f  X. \label{eq:j-var}
\eeq
\caca{The relevant scales are those larger than the binary separation (perturbations on smaller scales are affected in a complex way by the binary orbit and would require to be studied separately, as we discuss in Section \ref{sec:local_halo}). Using Eq.~\eqref{eq:a_ov_x}, we find that the dark matter mass corresponding to the binary scale when it decouples from the Hubble flow is of order $M_{\rm dm} \sim 0.1 M s_{\rm dec}$. As we will see below, the typical decoupling scale factor for binaries merging today is $s_{\rm dec} \sim 10^{-2} - 1$, so we conclude that the scales to be included in $\sigma_{\rm eq}$ in Eq.~\eqref{eq:j-var} are those corresponding to a dark matter mass larger than $\sim 10^{-3} M$.}

In principle the probability distribution for the total $\bs{j}$, which is the sum of two contributions (other PBHs and linear perturbations), can be computed by convolving the two probability distributions. This convolution is not analytic, however, so for simplicity we assume that for a given semi-major axis, the probability distribution of $j$ is given by Eq.~\eqref{eq:dP_dj}, with the characteristic value
\beq
j_X \approx 0.5 \left(1 + \sigma_{\rm eq}^2 /f^2 \right)^{1/2} X. \label{eq:jX}
\eeq

\subsection{Characteristic initial properties of binaries merging today}

\begin{figure}
\includegraphics[width = \columnwidth]{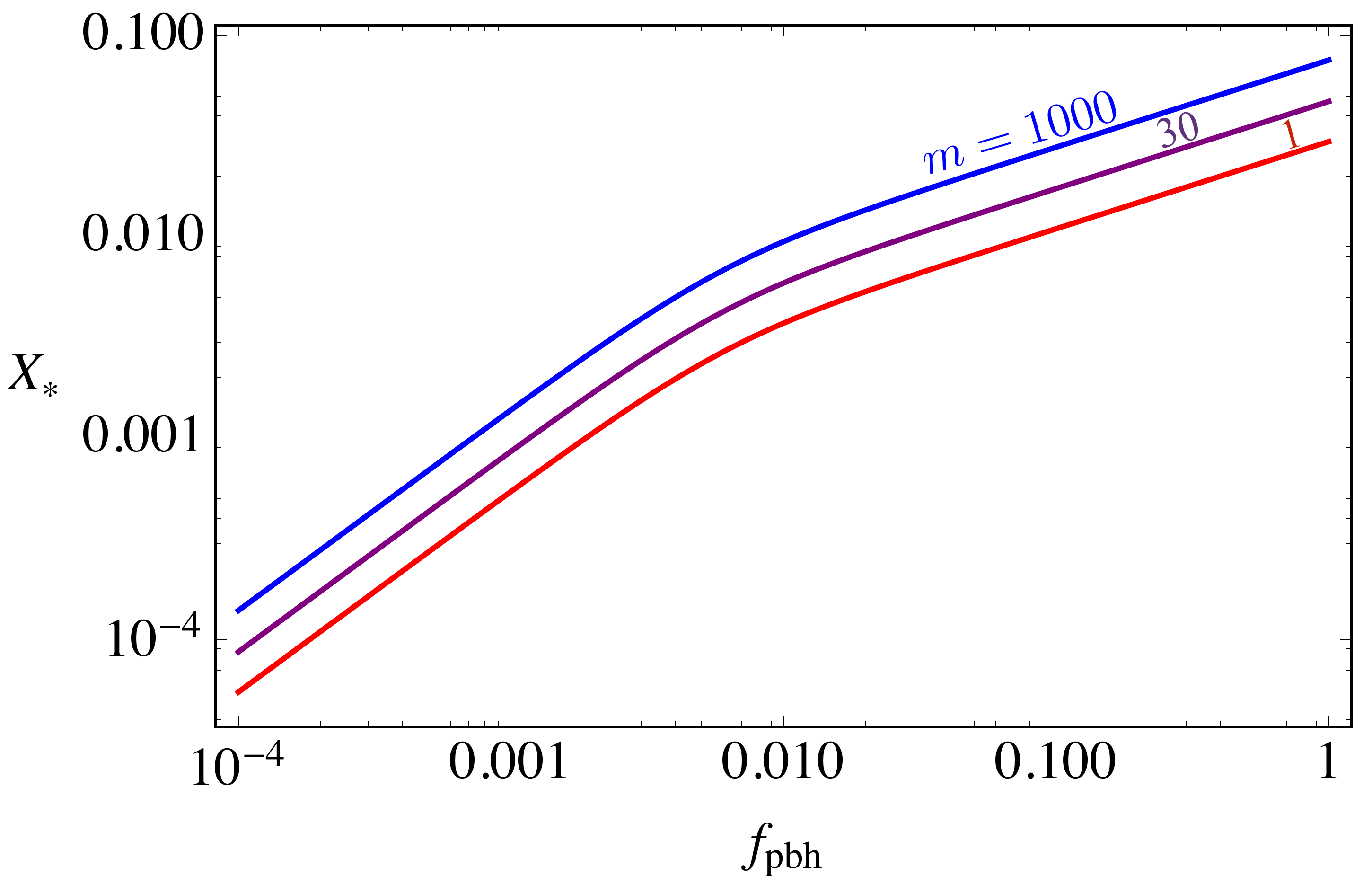}
\caption{Characteristic rescaled initial comoving separation $X \equiv (x/\overline{x})^3$ for PBH binaries that merge at the present time, as a function of the fraction of dark matter in PBHs. The curves are labeled by the PBH mass in units of $M_{\odot}$. We see that $X_* \ll 1$, indicating that PBH binaries merging today are rare pairs with initial separation much smaller than the characteristic inter-PBH separation. Here and in subsequent figures, the change of slope at $f \approx \sigma_{\rm eq} \approx 0.005$ is due to the change in the dominant tidal torque, from large-scale density perturbations at $f \lesssim \sigma_{\rm eq}$ to other PBHs at $f \gtrsim \sigma_{\rm eq}$.} \label{fig:Xstar}
\end{figure}

For initial eccentricities close to unity, i.e.~$j \ll 1$, which, as we will see shortly, is the relevant regime, the coalescence time through GW emission is given by \cite{Peters_64}
\barr
t &=& \frac{3}{170} \frac{a^4}{M^3} j^7. \label{eq:t_merge}
\earr
For a given $X$ hence $a$, there is a unique $j$ such that the merger time is $t$; using Eq.~\eqref{eq:a_ov_x}, it is given by
\beq
j(t; X) \equiv \left(\frac{170}3 \frac{t M^3 f^4}{(0.1 ~\overline{x})^4 X^{16/3}}\right)^{1/7}. \label{eq:j-t}
\eeq
The differential probability distribution of $(X, t)$ is then given by
\beq
\frac{d^2P}{dX dt} = \frac{dP}{dX} \frac{d P}{dt}\Big{|}_X = \frac{dP}{dX} \times \left[\frac{\partial j}{\partial t} \frac{d P}{dj}\Big{|}_X \right]_{j(t; X)}.
\eeq
The probability distribution of the rescaled nearest-neighbor separation is $dP/dX = \rme^{-X}$ \yah{(again, this assumes a random distribution of PBHs, and may take on different values in specific PBH formation models)}. Given that $j \propto t^{1/7}$, $\partial j/\partial t = j/(7 t)$. Using Eq.~\eqref{eq:dP_dj} we arrive at
\beq
\frac{d^2P}{dX dt} = \frac1{7 t} \rme^{-X} \mathcal{P}\left(\gamma_X\right),  \ \ \ \ \ \gamma_X \equiv \frac{j (t;X)}{j_X}.
\eeq
From Bayes' theorem, we obtain the probability distribution of $X$ for binaries merging after a time $t_0$:
\beq
\frac{dP}{dX} \Big{|}_{t_0} \propto \frac{d^2P}{dX dt}\Big{|}_{t_0}  \propto \rme^{-X} \mathcal{P}\left(\gamma_X\right) , \ \ t = t_0.
\eeq
We now seek the value $X_*$ for which this probability is maximized. We will see that $X_* \ll 1$, so we approximate $\rme^{-X} \approx 1$. We then need to solve
\beq
0 = \frac{\partial}{\partial X} \left[\frac{dP}{dX} \Big{|}_{t_0}\right]_{X_*} \propto \mathcal{P}'(\gamma_{X_*}) \frac{\partial \gamma_X}{\partial X}.
\eeq
Since $\gamma_X$ is strictly monotonic, this implies $\mathcal{P}'(\gamma_{X_*}) = 0$, which is achieved for $\gamma_{X_*} = \sqrt{2}$, i.e. 
\beq
j(t_0; X_*) = \sqrt{2} j_{X_*}.
\eeq
Solving for $X_*$, we obtain that the most probable value of $X$ for binaries merging today is
\beq
X_* \approx 0.032 ~f~ m^{5/37} (f^2 + \sigma_{\rm eq}^2)^{-21/74}.  \label{eq:Xstar}
\eeq
We show $X_*$ in Fig.~\ref{fig:Xstar}. We see that for all PBH masses and fractions of interest, $X_* \ll 1$, indicating that PBH binaries merging today are rare pairs with initial separation much smaller than the characteristic inter-PBH separation. This justifies our approximation to treat the effect of other PBHs as a perturbation on the nearly isolated binary.

\begin{figure}
\includegraphics[width = \columnwidth]{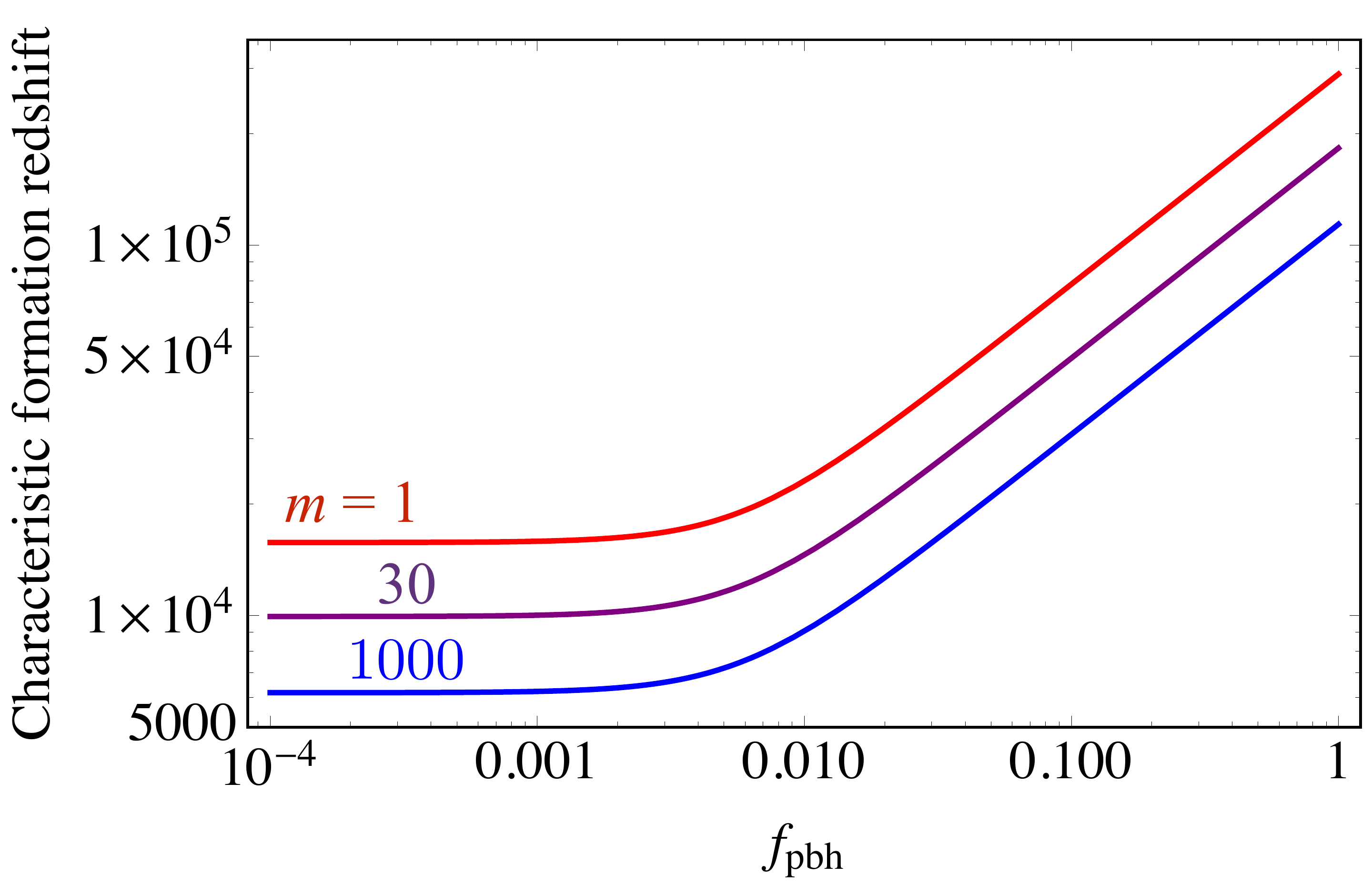}
\caption{Characteristic decoupling redshift of PBH binaries merging at the present time, as a function of the fraction of dark matter in PBHs. We see that PBH binaries typically form around matter-radiation equality for $f_{\rm pbh} \lesssim 0.01$, and much earlier for larger PBH fractions.} \label{fig:z_star}
\end{figure}

\begin{figure}
\includegraphics[width = \columnwidth]{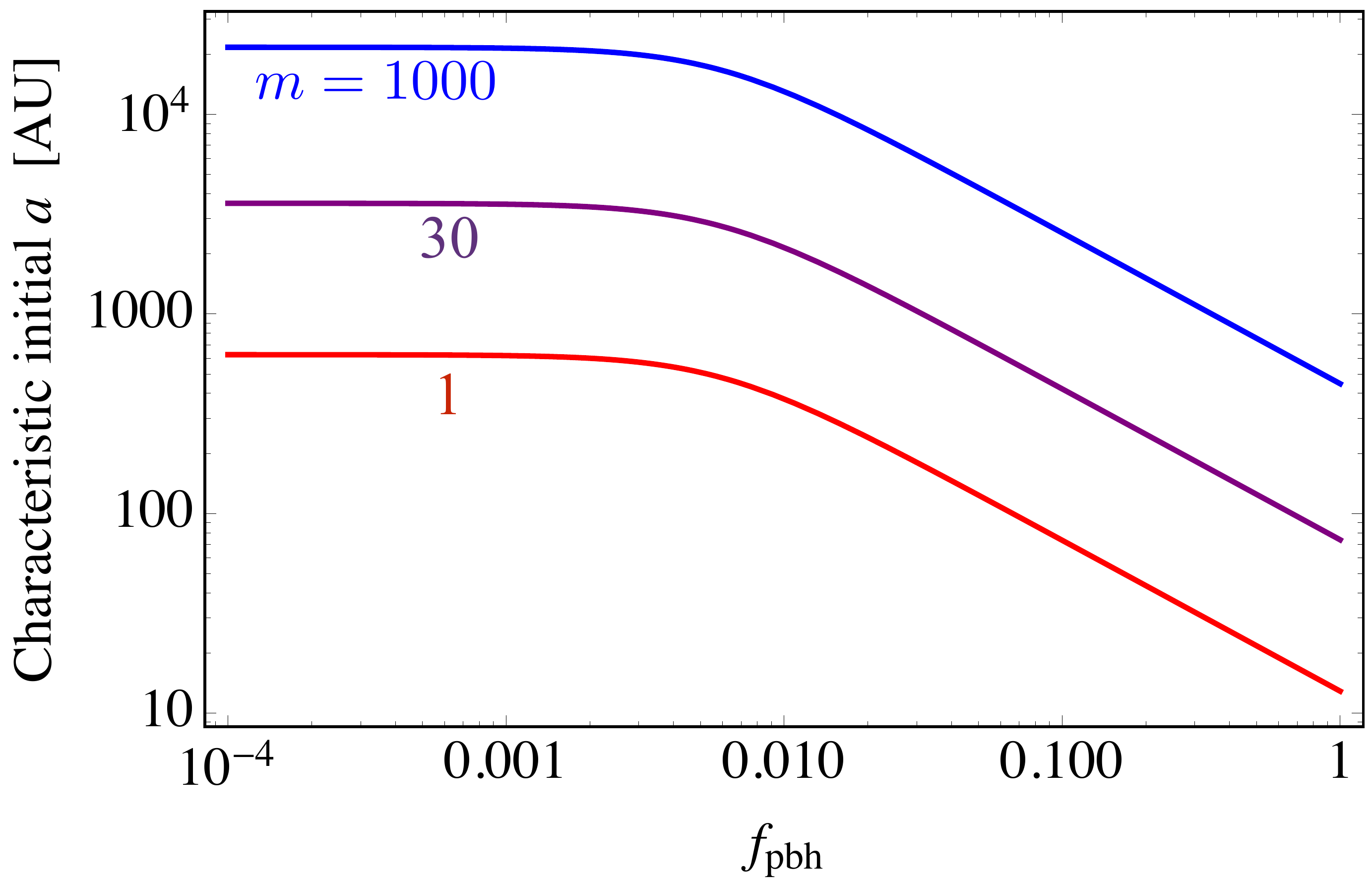}
\includegraphics[width = \columnwidth]{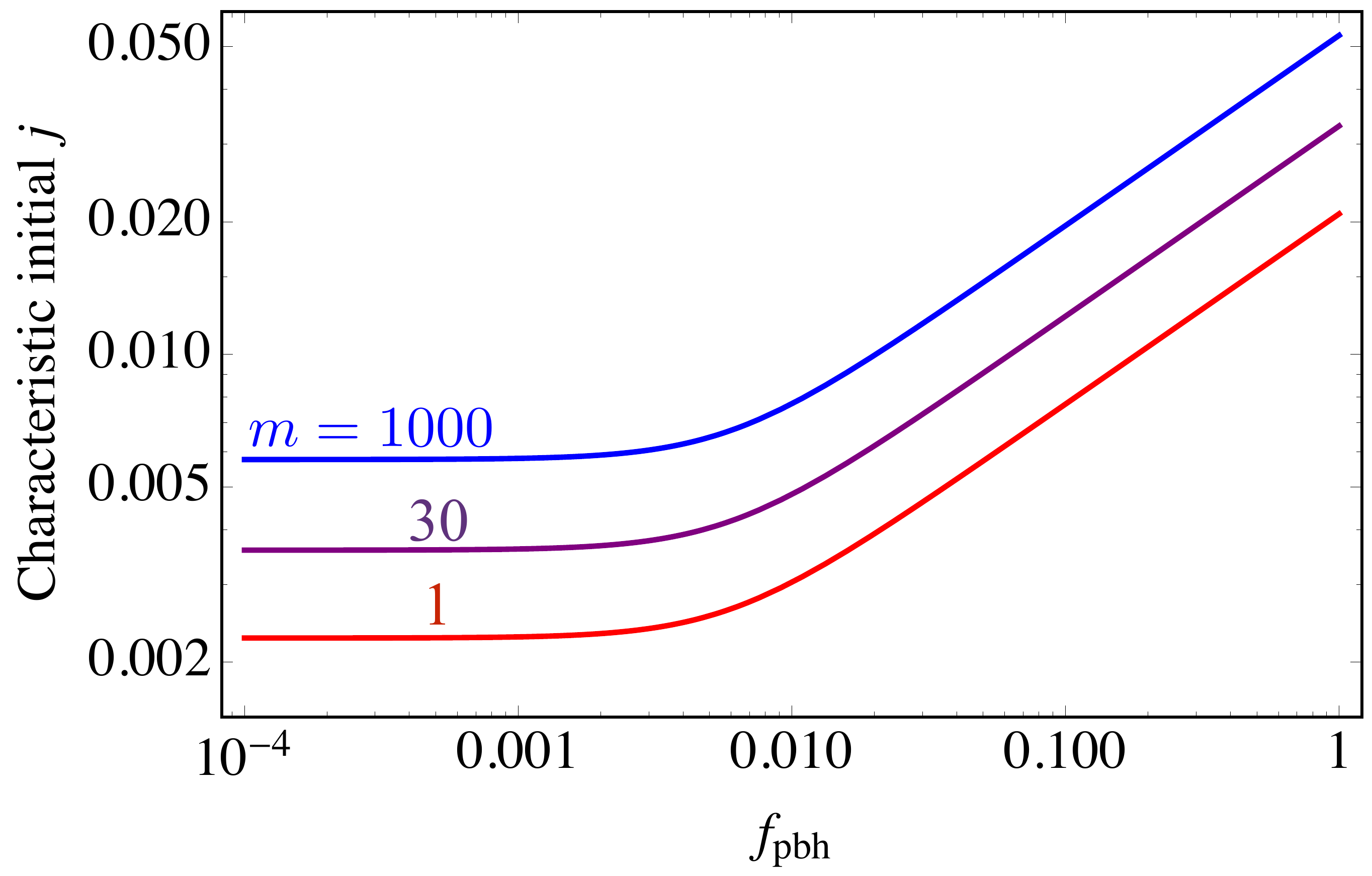}
\caption{Characteristic initial orbital elements (semi-major axis $a$ and reduced angular momentum $j = \sqrt{1-e^2}$) of PBH binaries merging at the present time.} \label{fig:aj_star}
\end{figure}

From our results in Sec.~\ref{sec:sma}, the characteristic redshift at which PBH binaries decouple from the Hubble flow is $z_* \approx 3 z_{\rm eq} /(X_*/f)$, which we show in Fig.~\ref{fig:z_star}. We find that all binaries merging today typically form prior to matter-radiation equality, and increasingly early for $f \gtrsim \sigma_{\rm eq}$. 
The characteristic semi-major axis $a_*$ is then obtained from Eq.~\eqref{eq:a_ov_x}, and the characteristic angular momentum $j_*$ is simply $j(t_0, X_*) = \sqrt{2} j_{X*}$, i.e., using Eq.~\eqref{eq:jX},
\yah{\barr
j_* &\approx& \frac1{\sqrt{2}} (\sigma_{\rm eq}^2 + f^2)^{1/2} (X_*/f) \nonumber\\
&\approx& 0.023 ~m^{5/37} (\sigma_{\rm eq}^2 + f^2)^{8/37}. \label{eq:jstar2}
\earr}
We show the characteristic initial orbital parameters in Fig.~\ref{fig:aj_star}.

\subsection{Merger rate}

We now have all the required ingredients to compute the merger rate. First of all, since the typical formation time is prior to matter radiation equality, the time of merger (i.e.~the value of coordinate time since the Big Bang) is approximately the time it takes to merge, for binaries merging today. The probability distribution of the time of merger is therefore
\beq
\frac{dP}{dt} = \int dX \frac{d^2P}{dX dt}= \frac1{7 t} \int dX \rme^{-X} \mathcal{P}(\gamma_X).
\eeq
Since the integrand peaks at $X_* \ll 1$, we may set $\rme^{-X} = 1$, and compute the integral analytically. Using $\gamma_X \propto X^{-37/21}$, and $\gamma_{X_*} = \sqrt{2}$, we find
\barr
\int dX \mathcal{P}(\gamma_X) &=& \frac{21}{37} \frac{X_*}{\sqrt{2}} \int d \gamma (\gamma/\sqrt{2})^{-58/37} \mathcal{P}(\gamma) \nonumber\\
&\approx& 0.59 ~ X_*.
\earr
The merger rate per unit volume at the present time $t_0$ is then obtained from
\beq
\frac{d N_{\rm merge}}{dt dV} = \frac12 f \frac{\rho_m^0}{M} \frac{d P}{dt}\Big{|}_{t_0} \approx 0.042~ X_* \frac{f \rho_m^0}{M t_0}, \label{eq:merger}
\eeq
where $\rho_m^0$ is the matter density at the present time, and the factor $1/2$ avoids double-counting of pairs .

\begin{figure}
\includegraphics[width = \columnwidth]{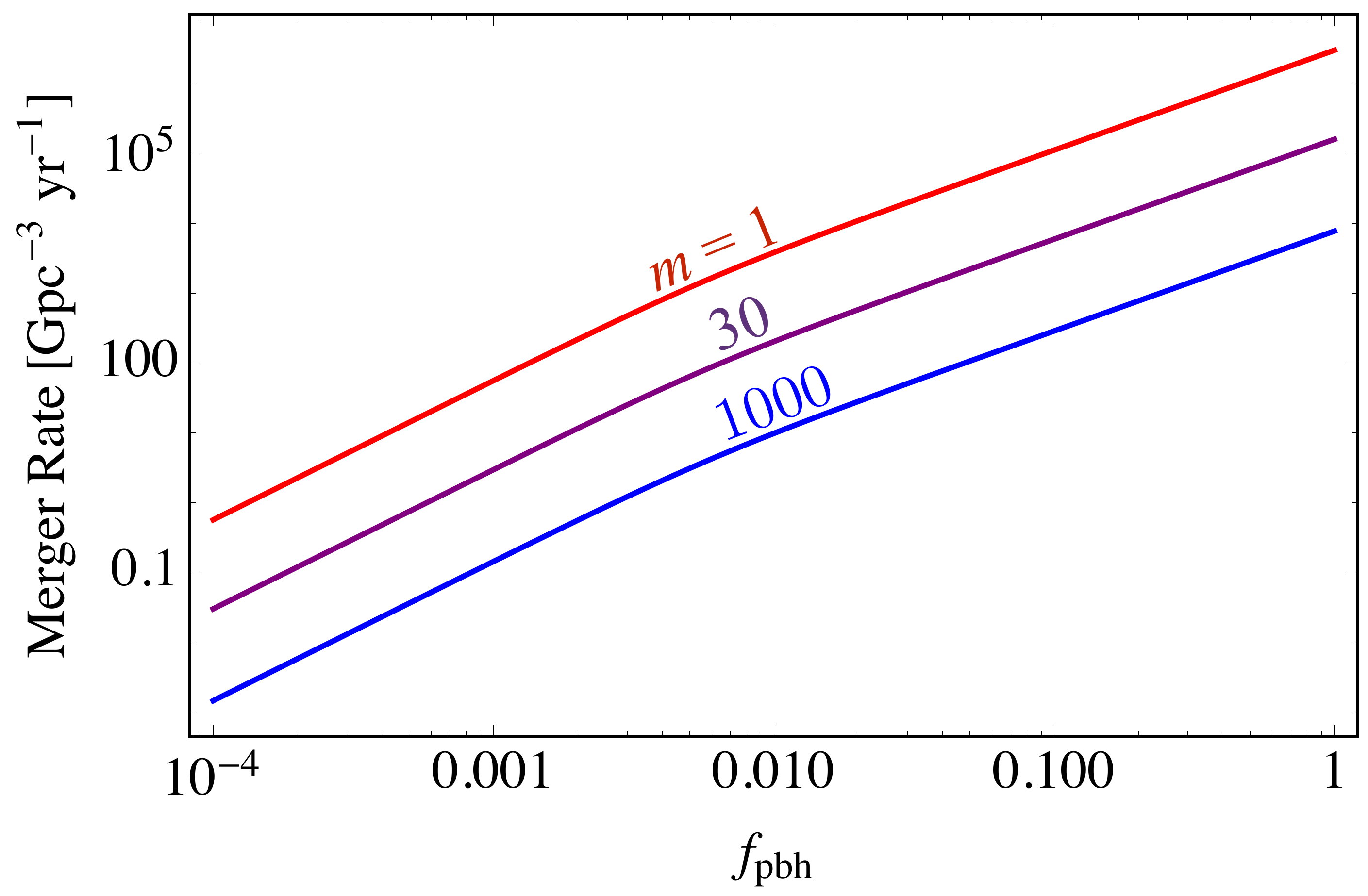}
\caption{PBH binary merger rate, as a function of PBH fraction $f_{\rm pbh}$ and mass $m = M/M_{\odot}$.} \label{fig:merger}
\end{figure}

We show the merger rate as a function of $f$ in Fig.~\ref{fig:merger}. It scales as $m^{-32/37} \approx m^{-0.86}$. For $f \gg \sigma_{\rm eq}$, it scales as $f^{53/37} \approx f^{1.41}$, and for $f \ll \sigma_{\rm eq}$ it scales as $f^2$. Note that this contrasts with the results of Ref.~\cite{Sasaki_16}, which did not account for torques by adiabatic density perturbations (i.e.~assumed $\sigma_{\rm eq} = 0$). In their case, the merger rate changes from $\propto f^{53/37}$ to $ \propto f^3$ at $f \lesssim 10^{-3}$, as PBH binaries typically form after matter-radiation equality in that case.

The next section is dedicated to check the most important assumption underlying this rate estimate, namely that between formation and merger, PBH binaries are mostly unaffected by their environment.

\section{Binary evolution between formation and merger} \label{sec:halos}

The goal of this section is to estimate the effect of interactions with the overall tidal field, other PBHs and baryons \emph{after} the binary has formed, once it is part of non-linear structures. 

\subsection{Purely gravitational interactions}

We begin by considering purely gravitational interactions of PBH binaries with dark matter, whether in the form of PBHs or otherwise. Before we start, let us point out that if PBHs do not make all of the dark matter, one must make assumptions about the rest of it. Given that the scales currently probed by CMB anisotropy and large-scale-structure measurements are \emph{significantly} larger than the scales of interest here, all bets are open regarding the appropriate model. For instance, the dark matter could be cold enough that its free streaming length is below current limts from Ly-$\alpha$ forest data \cite{Irsic_17}, yet be effectively warm on a scale containing a few PBHs. Similarly, the dark matter could be an ultra-light axion-like particle, massive enough to evade existing constraints \cite{Marsh_16}, yet light enough to have strong wavelike effects on the scales of interest. For definiteness, we shall assume that the rest of the dark matter is made of cold, collisionless particles with masses $\ll M$. In addition to being the simplest scenario, it is also that where the dark matter is expected to cluster the most, hence have the largest gravitational effects on PBH binaries. Making this assumption is therefore conservative. 

\subsubsection{Characteristic properties of early halos}

Consider a spherical region enclosing on average a total mass $M_h$. The number $N$ of PBHs it contains is Poisson distributed with mean $\langle N \rangle = f M_h/M$ and variance $\langle (\Delta N)^2 \rangle = \langle N \rangle$. For $\langle N\rangle \gg 1$, the distribution of perturbations on that mass scale is nearly Gaussian, with variance at equality
\barr
\sigma^2(M_h; \textrm{eq}) &\approx&  \sigma_{\rm eq}^2 + \frac{f^2}{\langle N\rangle}= \sigma_{\rm eq}^2 + f \frac{M}{M_h}. \label{eq:sigma_M}
\earr
\caca{Let us remark that the scales relevant to this section are typically larger than those relevant to the calculation of the torque in Sec.~\ref{sec:torques_dm} (if we consider only haloes containing at least a few PBHs), so the meaning of $\sigma_{\rm eq}$ is technically different in these two contexts. For simplicity, and assuming that the variance of perturbations varies slowly with mass on these scales, we may approximate them to be roughly equal.}

During the matter era, perturbations grow linearly with the scale factor, $\sigma(M_h, s) \approx s ~\sigma(M_h; \rm eq)$. Perturbations of mass scale $M_h $ typically collapse when $\sigma(M_h, s) \approx 1$, i.e.~at scale factor
\beq
s_{\rm coll}(M_h) \approx \left(\sigma_{\rm eq}^2 + f M/M_h\right)^{-1/2}. \label{eq:s_coll} 
\eeq
As a sanity check, with our assumed $\sigma_{\rm eq} = 0.005$, we find that the first small-scale structures form at $z \sim 20$ if $f = 0$, consistent with current estimates. 

Once a perturbation collapses and virializes into a halo, we assume its characteristic density $\rho_h$ is $\sim 200$ times the mean density at the time of collapse:
\beq
\rho_h \approx 200~ \overline{\rho}_m(s_{\rm coll}). \label{eq:rho_h1}
\eeq 
The variance of the relative velocity of two point masses in the halo is typically
\beq
v_h^2 \approx 2 \left( \frac{4 \pi \rho_h}{3} M_h^2 \right)^{1/3}. \label{eq:vh}
\eeq
The halo changes on a characteristic dynamical time
\beq
t_h \approx \sqrt{\frac{3}{4 \pi \rho_h}}. \label{eq:th}
\eeq
We are interested in a slightly different property, namely the characteristic halo mass a PBH binary (or any  mass element) is part of as a function of cosmic time. We shall make the simplest assumption, that at any given time most of the mass resides in halos that have just collapsed. In other words, inverting Eq.~\eqref{eq:s_coll}, we assume that the characteristic halo mass as a function of scale factor is
\beq
M_h(s) \approx f M s^2,\label{eq:Mh(s)}
\eeq
valid for $s \ll \sigma_{\rm eq}^{-1}$. At later times, standard adiabatic perturbations become larger than the small-scale Poisson contribution, and the characteristic mass $M_h(s)$ depends on the detailed functional form of $\sigma_{\rm eq}(M_h)$. Of course a PBH binary can only be part of a halo that contains at the very least 2 PBHs. For definiteness, we will consider halos as containing at least 10 PBHs. We define $s_{10}$ as the characteristic scale factor at which such halos (with mass $M_h = 10 M/f$) first form, 
\beq
s_{10} \equiv \left(\sigma_{\rm eq}^2 + f^2/10\right)^{-1/2}.\label{eq:s10}
\eeq 
To summarize, we take the following simple prescription: at a given scale factor $s \geq s_{10}$, a PBH binary is typically part of a halo of mass $f M s^2 = 10 M/f (s/s_{10})^2$, whose characteristic density, velocity dispersion and dynamical time are given by Eqs.~\eqref{eq:rho_h1}, \eqref{eq:vh} and \eqref{eq:th}.

\subsubsection{Relation of halo properties to characteristic PBH binary properties}

Here we write a few relations between the properties of the first halos and those of PBH binaries, which we will use repeatedly in the remainder of this section. We keep track of numerical factors in order not to add on to the uncertainty, but one should keep in mind that these are order-of-magnitude estimates.
%
We define the following dimensionless quantity
\beq
\epsilon \equiv \left(\frac{4 \pi \rho_h a^3}{3 M}\right)^{1/3} \equiv \epsilon_{10} (s_{10}/s), \label{eq:epsilon}
\eeq
where, using Eqs.~\eqref{eq:a_ov_x}, \eqref{eq:Xstar} and \eqref{eq:s10},
\barr
\epsilon_{10} &\approx& 0.6 ~(\sigma_{\rm eq}^2 + f^2/10)^{1/2} (X/f)^{4/3}\nonumber\\
&\approx& 0.006 ~ m^{20/111} \frac{(\sigma_{\rm eq}^2 + f^2/10)^{1/2}}{(\sigma_{\rm eq}^2 + f^2)^{14/37}} (X/X_*)^{4/3} \label{eq:epsilon10}
\earr
We see that $\epsilon_{10} \ll 1$ for all $(m, f)$ of interest. In words, PBH binaries are much ``denser" than the characteristic first halos, because they typically form much earlier on. 

An immediate consequence is that the orbital time of the binary is always much shorter than the dynamical time of the halo: 
\beq
\frac{t_{\rm orb}}{t_h} \sim \epsilon^{3/2} \ll 1. \label{eq:t_orb}
\eeq

\yah{Another relevant timescale is that of general-relativistic apsidal precession, which, for an equal-mass binary is (see e.g. Ref.~\cite{MTW})
\beq
t_{\rm prec} = \frac16 \frac{a}{M} j^2 t_{\rm orb}.
\eeq
For binaries merging at $t = t_0$, using Eq.~\eqref{eq:t_merge}, we may rewrite this as
\barr
t_{\rm prec} &=& \frac16 \left(\frac{170}{3} \frac{t_0}{M} j^{-7}\right)^{1/4} j^2 t_{\rm orb}\nonumber\\
 &\approx& 2 \times 10^5 m^{-1/4} j^{1/4} t_{\rm orb}.
\earr
In particular, for typical binaries with $j \approx j_*$, using Eq.~\eqref{eq:jstar2}, we get
\beq
\frac{t_{\rm prec}}{t_{\rm orb}} \approx 10^5 m^{-8/37} (\sigma_{\rm eq}^2 + f^2)^{2/37}.
\eeq
Using Eqs.~\eqref{eq:t_orb}, and \eqref{eq:epsilon}-\eqref{eq:epsilon10}, we therefore find, for the characteristic binaries merging today,
\beq
\frac{t_{\rm prec}}{t_h} \sim 10^5   m^{-8/37} (\sigma_{\rm eq}^2 + f^2)^{2/37} \epsilon_{10}^{3/2} (s_{10}/s)^{3/2}.
\eeq
Therefore we find that, until late times, the apsidal precession time is typically much longer than the characteristic dynamical time of the halo.} 

The ratio of the halo's pairwise velocity dispersion to a binary's circular velocity $v_{\rm bin} \equiv\sqrt{2 M/a}$ is 
\barr
\frac{v_h^2}{v_{\rm bin}^2} &\approx& \frac{a}{M} \left( \frac{4 \pi \rho_h}{3} M_h^2 \right)^{1/3} =(M_h/M)^{2/3} \epsilon.
\earr
For PBH fractions $f \gtrsim \sigma_{\rm eq}$, the mass of the first halos evolves according to Eq.~\eqref{eq:Mh(s)}, so we get
\barr
\frac{v_h}{v_{\rm bin}} &\approx& (10/f)^{1/3} \epsilon_{10}^{1/2} (s/s_{10})^{1/6}\nonumber\\
&\approx& 0.09 ~ m^{10/111} f^{-47/222}(s/s_{10})^{1/6} (X/X_*)^{2/3},
\earr
where the second line is valid for $f \gtrsim \sqrt{10} \sigma_{\rm eq} \sim 0.01$. For $f \gtrsim 0.01$, PBH binaries are therefore typically ``hard" relative to the first halos they are part of. Eventually, as the characteristic halo mass and virial velocity increase, this is no longer the case.

Finally, it is useful to rewrite the Hubble rate as follows: 
\beq
H = \sqrt{\frac{8 \pi}{3} \overline{\rho}_m} \approx \frac1{10} \sqrt{\frac{4 \pi \rho_h}{3}} \approx \frac{1}{10} t_h^{-1},
\eeq
and integrals over time become
\beq
\int \frac{dt}{t_h} = \int \frac{d\ln s}{H t_h} \approx 10 \int d \ln s. \label{eq:t_int}
\eeq
We now discuss how the orbital elements of PBH binaries may be affected by non-linear structure. We first consider the effect of the tidal field from the smooth halo, and then consider discrete interactions with other PBHs. 

\subsubsection{Torques by the tidal field from the smooth halo}

The rate of change of a binary's specific energy due to a tidal field $T_{ij}$ is $T_{ij} r_i \dot{r}_j = \frac12 T_{ij} \frac{d}{dt}(r_i r_j)$, where $r_i(t)$ is the binary's separation. Integrating over time, we see that the binary's energy hence semi-major axis are secularly conserved if $T_{ij}$ changes on a timescale much longer than the binary's period. We have shown in the previous section that $t_h \gg t_{\rm orb}$, and as a consequence, the halo's tidal field does not affect the semi-major axis of typical PBH binaries.

The tidal field exerts a torque per reduced mass 
\beq
\dot{\ell}_i = \epsilon_{ijk} r_j T_{kl} r_l. 
\eeq
Since $t_h \gg t_{\rm orb}$, we may average this equation over an orbital period:
\beq
\langle \dot{\ell}_i \rangle = \epsilon_{ijk}T_{kl} \langle r_j r_l \rangle \sim |T_{ij}| a^2.
\eeq
If the tidal field evolved on a timescale longer than the apsidal precession time, further averaging over the precession time would lead to $\dot{\ell}_i \propto \epsilon_{ijk}T_{kl} (\delta_{jl} - \hat{\ell}_j \hat{\ell}_l)$, i.e. $\dot{\bs{\ell}} \propto \bs{\ell} \times (\mathbf{T} \cdot \bs{\ell}) \ \bot \ \bs{\ell}$, in which case the magnitude of the angular momentum would also be secularly conserved. However, we saw in the previous section that $t_{\rm prec} \gg t_h$, so the tidal field of the halo can indeed change $\ell$. 

The characteristic tidal field $T_{ij} = - \partial_i \partial_j \phi$ is of order $|T_{ij}| \sim (4 \pi/3) \rho_h$, as can be seen from Poisson's equation. The torque on a binary with semi-major axis $a$ is therefore of order $\dot{\ell} \sim (4 \pi/3) \rho_h a^2$, implying, since $\dot{a} = 0$,
\beq
\frac{dj}{dt} \sim \frac{4 \pi}{3} \rho_h \frac{a^{3/2}}{M^{1/2}}.
\eeq
The tidal field of the smooth halo is roughly constant on a dynamical time $t_h$. During that time, the magnitude of the angular momentum changes by a amount $\Delta j_h$ of order
\beq
\Delta j_h \sim \frac{4 \pi}{3} \rho_h t_h  \frac{a^{3/2}}{M^{1/2}} \sim \epsilon^{3/2} \ll 1.
\eeq
After a time $\sim t_h$, the principal axes of the tidal field change direction, so the changes $\Delta j_h$ are uncorrelated from one dynamical time to the next. The angular momentum therefore undergoes a random walk on timescales $t \gg t_h$, and its variance grows as  
\barr
\langle \Delta j^2 \rangle &\sim& \int \frac{dt}{t_h} (\Delta j_h)^2 \sim \int \frac{dt}{t_h} \epsilon^3 \nonumber\\
&\sim& 10 \int d \ln s~ \epsilon^3 \sim 10~ \epsilon_{10}^3, \label{eq:Dj2}
\earr
where we have used Eq.~\eqref{eq:t_int}, and $\epsilon = \epsilon_{10}(s_{10}/s)$. \yah{Using Eqs.~\eqref{eq:epsilon10} and \eqref{eq:jstar2}, we find, for typical binaries merging today,
\barr
\frac{\langle \Delta j^2 \rangle}{j_*^2} &\sim& 0.004~ m^{10/37} \frac{(\sigma_{\rm eq}^2 + f^2/10)^{3/2}}{(\sigma_{\rm eq}^2 + f^2)^{58/37}} \ll 1.
\earr
}
We therefore conclude that the overall halo's tidal field does not significantly alter the orbital parameters of PBH binaries merging today.

\subsubsection{Distant encounters with other PBHs}

Let us now consider discrete interactions with other PBHs. We start by considering distant encounters, i.e.~those whose distance of closest approach $r_p$ is at least a few times $a$, so that their interaction is tidal.

Consider a single PBH approaching a PBH binary on a hyperbolic orbit with impact parameter $b$ and relative velocity at infinity $v$. Conservation of energy and angular momentum imply the following relations between $(b, v)$ and the distance of closest approach $r_p$:
\barr
b^2 &=& r_p^2 + \frac{6 M r_p}{v^2}, \label{eq:b-r_p}\\
r_p &=& \frac{b}{[1 + (3 M /b v^2)^2]^{1/2} + 3 M/bv^2}.
\earr 
By conservation of angular momentum, the velocity at pericenter is $v_p = \frac{b v}{r_p}$. The characteristic interaction time is therefore 
\beq
t_p \sim \frac{r_p}{v_p} = \frac{r_p^2}{bv}.
\eeq
The change in specific angular momentum is of order 
\barr
\Delta \ell \sim \frac{M}{r_p^3} a^2 t_p \sim \frac{M a^2}{r_p b v}.  
\earr
The change of specific energy $E/M$ is at most (provided the interaction is prompt, i.e.~$t_p \lesssim t_{\rm orb}$) of order
\beq
\frac{\Delta E}{M} \sim \frac{M}{r_p^3} a \sqrt{\frac{M}{a}}  t_p \sim \frac{M^{3/2} a^{1/2}}{r_p bv},
\eeq
hence the fractional change in semi-major axis is at most
\beq
\frac{\Delta a}{a} = \frac{\Delta E}{E} \sim \frac{M^{1/2} a^{3/2}}{r_p b v} .
\eeq
The change in $j$ is of the same order:
\beq
\Delta j = \frac{\Delta \ell}{\sqrt{2 M a}} - \frac12 j \frac{\Delta a}{a} \sim \frac{\Delta \ell}{\sqrt{Ma}} \sim \frac{M^{1/2} a^{3/2}}{r_p b v}.
\eeq
Note that limiting ourselves to tidal distant encounters implies that these changes are always small: using Eq.~\eqref{eq:b-r_p}, with $r_p \gg a$, we get
\beq
\frac{\Delta a}{a} \lesssim \Delta j \ll \left(1 + a v^2/6 M\right)^{-1/2} \leq 1.
\eeq
The merger timescale can only be significantly affected if $a$ or $j$ change by a fractional amount of order unity. While $\Delta a/a \ll 1$ for distant encounters, since the characteristic $j$ is small, we do have to make sure that $\Delta j \ll j$. 

To change $j$ by an amount greater than $\Delta j_0$ requires 
\beq
r_p b \lesssim \frac{M^{1/2} a^{3/2}}{v \Delta j_0}. 
\eeq
Defining $\alpha \equiv b v^2 /(3 M)$ and $v_a \equiv \sqrt{M/a}$, this condition implies
\beq
\frac{\alpha^3}{1 + \sqrt{1 + \alpha^2}} \lesssim \frac{v^3}{v_a^3 \Delta j_0}.
\eeq
This has the approximate solution
\beq
\alpha^2 \lesssim \alpha_0^2 \equiv \left(\frac{v^3}{v_a^3 \Delta j_0}\right)^{2/3} + \frac{v^3}{v_a^3 \Delta j_0}. \label{eq:alpha0}
\eeq
From this we obtain the cross section to change $j$ by more than $\Delta j_0$:
\barr
\sigma_{\Delta j \geq \Delta j_0} (v)&=& \pi \left(\frac{3 M}{v^2}\right)^2 \alpha_0^2 \nonumber\\
&\sim& \frac{M a}{v^2} \frac{1}{\Delta j_0^{2/3}} + \frac{M^{1/2} a^{3/2}}{v} \frac1{\Delta j_0}. \label{eq:sigma_Dj}
\earr
The first term in this cross section corresponds to the limit of nearly parabolic encounters, and the second term to the limit of quasi-straight line trajectories. The former has the same dependence as equation (19) of \cite{Heggie_96} (once translated from $e$ to $j$), where this cross section was computed exactly in the quasi-parabolic-orbit limit. The numerical prefactor found in Ref.~\cite{Heggie_96} for equal-mass objects is $\approx 7$.

The number of encounters that change $j$ by more than $\Delta j_0$, per binary, is then 
\beq
N_{\Delta j \geq \Delta j_0} = \int dt \frac{f \rho_h}{M} v_h \sigma_{\Delta j \geq \Delta j_0}(v_h),
\eeq
where $f \rho_h/M$ is the number density of PBHs, that is,
\barr
&&N_{\Delta j \geq \Delta j_0} \sim f \left[\frac{a}{\Delta j_0^{2/3}} \int dt \frac{\rho_h}{v_h} +   \frac{(a^3/M)^{1/2}}{\Delta j_0} \int dt \rho_h\right]\nonumber\\
&\sim& f \left[ \frac1{\Delta j_0^{2/3}} \int \frac{dt}{t_h} \epsilon \left(\frac{M}{M_h}\right)^{1/3} + \frac1{\Delta j_0} \int \frac{dt}{t_h} \epsilon^{3/2} \right] \nonumber\\
&\sim& 10 f \left[ (f/10)^{1/3} \left(\frac{\epsilon_{10}^3}{\Delta j_0^2}\right)^{1/3}  + \left(\frac{\epsilon_{10}^{3}}{\Delta j_0^2}\right)^{1/2} \right].  
\earr
An initially narrow distribution in $j$ is broadened by impulsive torques from other PBHs up to the characteristic $\Delta j_0$ for which $N_{\Delta j \geq \Delta j_0} \approx 1$, which is approximately
\barr
\Delta j_0^2 \sim \epsilon_{10}^3 \times \max \left[(f/10) (10 f)^3, (10 f)^2 \right]  \sim (10 f)^2 ~\epsilon_{10}^3.~~~~~ \label{eq:Dj2-impulsive}
\earr
This is a factor of $\sim 10 f^2$ larger than Eq.~\eqref{eq:Dj2}, and is therefore still small relative to $j_*^2$, even for $f \approx 1$, and even if multiplied by a factor of $\sim 10$. Therefore we conclude that distant encounters with other PBHs do not significantly affect the orbital parameters of PBH binaries merging today.

%

\subsubsection{Close encounters with other PBHs}

Let us estimate the probability of close encounters, i.e.~those with closest approach $r_p \lesssim a$. The cross section from such encounters is obtained from Eq.~\eqref{eq:b-r_p}:
\beq
\sigma_{\rm close}(v) \sim \pi \left[ a^2 + 2 \frac{M a}{v^2} \right].
\eeq
The first term is just the geometric cross section and the second term accounts for gravitational focusing. Therefore the total number of such encounters per binary is
\barr
N_{\rm close} &=& \frac{f}{M} \int dt \rho_h v_h \sigma_{\rm close}(v_h) \nonumber\\
&\sim& f \left[ \frac{a^2}{M} \int dt \rho_h v_h +  a \int dt \frac{\rho_h}{v_h} \right]. 
\earr
While the second term is dominated by the first halos as $v_h$ increases with time, the first term also gets a contribution from low redshifts. Indeed, at $z \lesssim 20$, the characteristic halo mass increases exponentially with redshift due to the logarithmic behavior of $\sigma(M_h)$, and so does $v_h \sim M_h^{1/3} \rho_h^{1/6}$. 

Let us start by evaluating the low-redshift contribution, i.e. the number of close encounters in Milky-Way-like halos, with characteristic density $\rho_{\rm MW} \approx 200 \rho_m^0$ and velocity dispersion $v_{\rm MW} \approx 200$ km/s:
\barr
\frac{N_{\rm close}}{dt}\Big{|}_{\rm mw} &\sim& f\frac{a^2}{M} \rho_{\rm MW} v_{\rm MW} \nonumber\\
&\sim& 10^{-4} f \left(\frac{a}{10^4 \textrm{AU}}\right)^2 \frac{30 M_{\odot}}{M} ~\textrm{Gyr}^{-1}.
\earr
The probability of close encounters in Milky-Way-like halos is therefore clearly negligible. 

The contribution from the first halos is dominated by the gravitational focusing term, since $v_h \gg v_{\rm orb}$ in the first halos:
\beq
 N_{\rm close} \sim \int \frac{dt}{t_h} \epsilon \left(\frac{M}{M_h}\right)^{1/3} \sim 10 f (f/10)^{1/3} \epsilon_{10}. 
 \eeq
From Eq.~\eqref{eq:epsilon10} we see that this number is at most $\sim 0.01$ for $f \sim 1$, and very small for $f \ll 1$. 

We therefore conclude that close encounters are unlikely, and as a consequence we need not worry about their detailed consequences. Let us point out, as additional reassurance, that since PBH binaries are typically hard relative to the first halos, they can not be disrupted (``ionized"), as the perturbing PBHs do not have sufficient energy to do so.

\subsubsection{Dynamical friction by dark matter particles}

Ref.~\cite{Hayasaki_16} pointed out that if PBHs are a subdominant component of dark matter, and if the bulk of dark matter is made of particles with masses $\ll M$, they would exert dynamical friction on PBH binaries, possibly making them merge in a short timescale. Here we revisit this process using our analytic estimates for the properties of the first halos.

The rate of hardening of a binary in a background of point masses was estimated in Ref.~\cite{Quinlan_96}. The effect is largest for hard binaries. In the limit of hard, highly-eccentric binaries, the results of \cite{Quinlan_96} are approximately 
\beq
\frac{d \ln a}{dt} \approx -30 \frac{\rho_h a}{v_h} \approx -\frac{3}{t_h} \epsilon (M/M_h)^{1/3}.
\eeq
Therefore we see that $|\Delta \ln a| \sim 30 \epsilon_{10} \ll 1$.

The fitting function for the rate of change of eccentricity given in Ref.~\cite{Quinlan_96} translates to
\beq
\frac{d \ln j}{d \ln a} \approx e^2(k_1 + k_2 e) j^{2 (k_0 -1)} \approx (k_1 + k_2) j^{2(k_0 -1)}, 
\eeq
where the constants $k_i$ are of order unity and depend on $v_h/v_{\rm bin}$, and the second approximation holds for $e \approx 1$. The minimum value of $k_0$ estimated by Ref.~\cite{Quinlan_96} is $k_0 \sim 0.75$. Therefore we find that $\Delta \ln j$ is at most $\sim 10 \epsilon_{10}/j_*^{1/2}  \sim 10 (\sigma_{\rm eq}^2 + f^2/10)^{1/2}/(\sigma_{\rm eq}^2 + f^2)^{1/4} (X_*/f)^{5/6} \ll 1$.

From these analytic estimates, we conclude that dynamical friction by dark-matter particles does not significantly affect the orbital parameters of PBH binaries merging today. 

\subsubsection{A note on local particle-dark-matter halos}\label{sec:local_halo}

The previous paragraphs focused on large halos, containing a few PBHs on average. In particular, for $\sigma_{\rm eq} \lesssim f \ll 1$, such halos only collapse at $s_{10} \sim \sqrt{10}/f$, i.e.~significantly after matter-radiation equality.

In addition to being part of large halos, PBHs get ``clothed" in a local halo of bound dark matter particles if $f_{\rm pbh} < 1$ \cite{Mack_07, Ricotti_07, Eroshenko_16b}. While accretion onto a binary must differ from accretion onto a point mass, we nevertheless generically expect that a dark-matter mass of order $M_{\rm dm} \sim M$ becomes bound to the binary by matter-radiation equality. \caca{If the bulk of the dark matter is a thermal relic with a weak-scale annihilation cross section, the local dark matter halo could moreover be a luminous source of annihilation products \cite{Lacki_10}.} It is difficult to estimate analytically the properties of such a local halo and the effect it may have on PBH binaries (see Ref.~\cite{Hayasaki_16} for an attempt to do so). It is however a well-posed numerical problem, the study of which we defer to future work. \\[5pt]

To conclude this section, we find that the orbital parameters of PBH binaries formed in the early Universe are not significantly affected by halos large enough to contain $\sim 10$ PBHs. However, we emphasize that this conclusion relies on very simple analytic estimates for complex dynamics. \yah{For instance, one of our underlying assumptions is that each generation of halos gets mostly tidally stripped as they get engulfed in the next generation of more massive halos. It might be that most of the first halos actually survive as substructure of the subsequent generation, in which case the phase-space parameters determining collision rates would significantly differ from our estimates. A definitive answer to the problem of survival of PBH binaries in early halos is therefore likely to require more detailed numerical studies.}

\subsection{Effect of baryon accretion}

Ref.~\cite{Hayasaki_16} suggested that a thin circumbinary accretion disk may form around PBH binaries once the Bondi radius exceeds the binary separation. Such a disk would then exchange energy and angular momentum with the binary \cite{Hayasaki_09}. While it is not clear what kind of accretion flow would take place (thin, slim, thick disk or advection-dominated flow, see e.g.~\cite{Abramowicz_13} for a review), here we attempt to give an estimate of the magnitude of this effect.

\subsubsection{Estimate of the work and torque on the binary}

Let us first consider a quasi-spherical accretion flow onto the binary. Assuming the binary separation is less than the Bondi radius, the characteristic velocity of the gas at the orbital separation is of order the free fall velocity, which is of the order of the orbital velocity $v \sim \sqrt{M/a}$. The drag force on each black hole is of the order $\dot{M} v$, i.e., if we assume the Bondi-Hoyle-Lyttleton accretion rate locally onto each black hole, \cite{Hoyle_39, Bondi_44, Bondi_52},  
\beq
M \dot{v} \sim - 4 \pi \rho_{\rm gas}^{(a)} \frac{M^2}{v^3} v \sim - 4 \pi \rho_{\rm gas}^{(a)} M a,  
\eeq
where $\rho_{\rm gas}^{(a)}$ is the local gas density. Now, the total accretion rate on the binary is such that $\dot{M} = 4 \pi \rho_{\rm gas} r^2 v_r =$ constant, so the gas density at the binary's orbit is of order $4 \pi \rho_{\rm gas}^{(a)} \sim \dot{M}/(a^2 \sqrt{M/a})$. Hence we get
\beq
M \dot{v} \sim - \dot{M} \sqrt{M/a}.
\eeq
The binary loses energy at a rate $\dot{E} \sim M \dot{v} v$ and angular momentum at a rate $\dot{L} \sim M \dot{v} a$, i.e., denoting the binary's total mass by $M_{\rm bin} = 2 M$, we get
\barr
\dot{E} &=& - A~  \dot{M} \frac{M_{\rm bin}}{a}, \label{eq:Edot-Mdot} \\
\dot{L} &=& -B~ \dot{M} \sqrt{M_{\rm bin} a},\label{eq:Ldot-Mdot}
\earr
where $A$ and $B$ are dimensionless numbers of order unity.

Let us now consider instead a thin circumbinary accretion disk with surface density $\Sigma$ and effective shear viscosity $\nu$. The disk is truncated at an inner edge $r_{\rm in}$, where viscous torques balance gravitational torques \cite{Artymowicz_94}. This inner edge is typically around one of the lowest Lindblad resonances \cite{Goldreich_79}, i.e. $r_{\rm in} \sim 2 a$, as confirmed in numerical simulations \cite{Artymowicz_94} with typical Shakura-Sunyaev \cite{Shakura_73} viscosity parameters. The torque $\dot{L}$ on the binary is the opposite of the torque on the disk, which is approximately the viscous torque at the inner edge, i.e. \cite{Artymowicz_94}
\beq
\dot{L} \sim - 3 \pi \nu \Sigma \sqrt{M_{\rm bin} r_{\rm in}}.
\eeq
The effective viscosity also determines the radial inflow, hence the accretion rate \cite{Pringle_81}:
\beq
\dot{M} \sim 3 \pi \nu \Sigma.
\eeq
Dividing these two equations, we obtain again Eq.~\eqref{eq:Ldot-Mdot}. Of course, the parameter $B$ depends on the binary's eccentricity and on the details of the accretion physics, but should be of order unity (see also Ref.~\cite{Hayasaki_13} for a similar parametrization).

The scaling \eqref{eq:Ldot-Mdot} is confirmed in numerical simulations of thin circumbinary disks. Ref.~\cite{MacFadyen_08} simulated a circumbinary accretion disk around a circular binary. Dividing their Eqs.~(18) and (20) gives our Eq.~\eqref{eq:Ldot-Mdot} with $B \approx 2.4$. While \cite{MacFadyen_08} only explicitly resolved the region $r \geq a$, the more recent simulations of Ref.~\cite{Tang_17} resolves the interior region $r \leq a$ and the individual accretion ``minidisks" around each black hole. They find that the dominant contribution of the torque arises from gas streams close to the individual black holes. Recasting their equation (14) in physical units, and setting the sink timescale to the characteristic viscous timescale, translates again to our Eq.~\eqref{eq:Ldot-Mdot} with $B \approx 5.6$.

The rate of energy change can be obtained from the following consideration. For a perturbing potential $\Phi$ of the form $\Phi \propto \phi_{m l}(r) \rme^{i (m \theta - l \Omega_b t)}$, where $\theta$ is the polar angle and $\Omega_b \equiv \sqrt{M_{\rm bin}/a^3}$, the combination $E - \frac{l}{m} \Omega_b L$ is constant. \yah{Provided the torque} is dominated by low-order $(m, l)$, we therefore find $\dot{E} \sim \Omega_b \dot{L}$, which translates to Eq.~\eqref{eq:Edot-Mdot}. Here again, the parameter $A$ depends on eccentricity and the details of the accretion disks, but should be of order unity.

Therefore, we expect the simple relations \eqref{eq:Edot-Mdot} and \eqref{eq:Ldot-Mdot} to hold under rather general circumstances. \yah{From these relations, and using $E = -\frac18 M_{\rm bin}^2/a$ and $L = \frac14 M_{\rm bin}^{3/2} a^{1/2} j$ for an equal-mass binary, it is straightforward to show that
%
\beq
\frac{\dot{a}}{a}  \sim  \frac{d j}{dt}  \sim - \frac{\dot{M}}{M}, \label{eq:adot-Mdot}
\eeq
with uncertain numerical prefactors. Note, that the sign of the effect itself is uncertain: the binary could instead extract energy and angular momentum from the accretion flow.}

\subsubsection{Accretion rate and total change in orbital parameters}
 
Our next step is now to estimate the accretion rate $\dot{M}$. Accretion is typically highly time-dependent and may proceed in bursts (see e.g.~Fig.~7 of Ref.~\cite{MacFadyen_08}), whose amplitude and timescale cannot be simply estimated. However, provided the Bondi radius is larger than the binary separation, the large-scale accretion flow should be roughly of the Bondi-Hoyle-Lyttleton type \cite{Hoyle_39, Bondi_44, Bondi_52}: outside the sonic radius there should be little difference between accretion onto a point mass or a binary\footnote{We thank Geoffrey Ryan for pointing this out.}. Therefore, on timescales long compared to the Bondi time, the average accretion rate ought to be close to the Bondi value, $\langle \dot{M} \rangle \approx \dot{M}_{\rm B}$. Since the total change in orbital parameters is proportional to $\int dt \dot{M}/M$, it is this long-timescale accretion rate that is relevant, rather than the possibly large short-time fluctuations. We denote by $\dot{m} \equiv \dot{M}/\dot{M}_{\rm Edd}$, where\footnote{Our convention follows that of Refs.~\cite{Ricotti_08, YAH_17} but differs from the accretion-disk litterature where $\dot{M}_{\rm Edd}$ is often defined with a factor of 10 larger.} \caca{$\dot{M}_{\rm Edd} c^2 \equiv L_{\rm Edd} \approx 2~ M c^2$ Gyr$^{-1}$ is the Eddington luminosity. We therefore have
\beq
\int dt \frac{\dot{M}}{M} \sim 2 \int \frac{\dot{m}~dt}{\textrm{Gyr}}. 
\eeq}
The rate of accretion from the background baryon gas was computed in Ref.~\cite{YAH_17} accounting for relative motions of baryons and PBHs and Compton cooling and heating. At redshifts less than a few hundred, which dominate the integral, the characteristic dimensionless accretion rate they find is $\dot{m} \sim 10^{-5} (M/M_{\odot})$. Therefore, we get
\beq
\int^{t_{\max}} dt \frac{\dot{M}}{M} \sim 2 \times 10^{-5} \frac{t_{\max}}{\textrm{Gyr}} \frac{M}{M_{\odot}}.
\eeq
At low enough redshift, once the binaries are part of large halos, we expect the accretion to be cutoff due to the large non-linear velocities and heating of the gas \cite{Ricotti_08}. Cutting off the integration at $z \approx 10$ corresponds to $t_{\max} \sim 0.5$ Gyr, so we get
\beq
\int dt \frac{\dot{M}}{M} \sim 10^{-5}  \frac{M}{M_{\odot}},
\eeq
which matches what Ref.~\cite{Hayasaki_16} estimated from the results of Ref.~\cite{Ricotti_08}. Therefore we conclude that, even if the coefficient in Eq.~\eqref{eq:adot-Mdot} is $\sim 10-100$, the semi-major axis should not be significantly affected by an accretion disk. This contrasts with the results of Ref.~\cite{Hayasaki_16} who found an orbital decay timescale much shorter than the Hubble time. The difference can be traced back to their estimate of the characteristic disk mass $M_{\rm cbd} \equiv \pi a^2 \Sigma$ by $M_{\rm cbd} \sim H^{-1}\dot{M}$, instead of the more appropriate order of magnitude $M_{\rm cbd} \sim t_{\rm visc} \dot{M}$. Ref.~\cite{Hayasaki_16} therefore seem to have overestimated the effect of the accretion disk by a factor $\sim H^{-1}/ t_{\rm visc} \gg 1$. Nevertheless, if the coefficient in Eq.~\eqref{eq:adot-Mdot} is large, and for large enough PBH masses, the change in $j$, while small in absolute value, could still exceed the characteristic initial value for PBH binaries merging today (see Fig.~\ref{fig:aj_star}). 

\yah{If this is the case, and accretion efficiently extracts angular momentum, binaries that would have otherwise merged today may merge much earlier on. In the extreme case where most binaries merge quickly, a high-redshift gravitational-wave background would result \cite{Hayasaki_16}. Conversely, if accretion tends to circularize eccentric binaries, they may merge on a much longer timescale. More generally, if accretion significantly affects orbital parameters, the probability distribution of merger times, hence the merger rate, could be drastically different from what we have estimated in Section \ref{sec:initial_cond}. This warrants further work, most likely numerical simulations, to investigate this issue in more detail.} 

%

\section{PBH binary formation in present-day halos} \label{sec:bird}

It was pointed out in Ref.~\cite{Bird_16} (hereafter, BCM) that PBH binaries can also form in present-day halos through gravitational bremsstrahlung: if two PBHs pass close enough to each other, they may radiate a sufficient amount of energy in gravitational waves to become bound. The binaries formed through this pathway are typically very tight and highly eccentric \cite{Cholis:2016kqi}, and coalesce within a timescale much shorter than a Hubble time, so that the merger rate is approximately equal to the capture rate. BCM found that the merger rate is dominated by the smallest halos, of a few hundred solar masses, and is of order $\sim 1$ Gpc$^{-3}$ yr$^{-1}$ if PBHs make all of the dark matter. This is significantly lower than the merger rate of binaries formed in the early Universe. However, the calculation of BCM did not account for the contribution of Poisson fluctuations to density perturbations when estimating the characteristic density and velocity dispersion of the smallest halos. Given how sensitively these depend on the variance of perturbations, it is worth revisiting this calculation. Throughout this section we assume $f = 1$.

\subsection{General considerations}

The cross section for two equal masses to become bound due to gravitational radiation is \cite{Quinlan_89}
\beq
 \sigma_{\rm gw}(v) = 4 \pi \left( \frac{85 \pi}3 \right)^{2/7} \frac{M^2}{v^{18/7}} \approx 45 \frac{M^2}{v^{18/7}}, 
\eeq
where $v$ is the relative velocity of the two PBHs at large separation. The rate of binary formation in a given halo is therefore
\barr
\Gamma &=& \frac12 \int d^3 r \frac{\rho(\bs{r})^2}{M^2} \langle v \sigma_{\rm gw} \rangle(\bs{r})
\sim 20 ~ M_h \rho_h v_h^{-11/7}. ~~~~
\earr
Using Eq.~\eqref{eq:vh} for virialized halos, we have $v_h \sim 2 M_h^{1/3} \rho_h^{1/6}$, so that 
\beq
\Gamma \sim 10 ~M_h^{10/21} \rho_h^{31/42}.
\eeq
A simple prescription for the characteristic halo density is that it is $\sim 200$ times the mean density at the time of collapse. Neglecting the effect of dark energy at low redshift, the characteristic redshift of collapse of perturbations of mass $M_h$ is  
\beq
z_{\rm coll} \sim \sigma(M_h), 
\eeq
where $\sigma(M_h)$ is the variance of linear perturbations on the mass scale $M_h$ extrapolated to the present time. Therefore, 
\beq
\rho_h \sim 200 ~[\sigma(M_h)]^3~ \overline{\rho}_m^0. \label{eq:rho_h}
\eeq
Note that this is consistent with the asymptotic behavior of the fitting formula of Ref.~\cite{Prada_12} for the mass-concentration relation: for NFW profiles, $\rho_h \sim 200 ~\overline{\rho}_m^0 ~ c^3$, and the concentration $c$ scales nearly linearly with $\sigma(M_h)$ at large values. This implies
\beq
\Gamma \sim 450 ~ \left(\overline{\rho}_m^0\right)^{31/42} M_h^{10/21}[\sigma(M_h)]^{31/14}. \label{eq:Gamma_h}
\eeq
The merger rate per unit volume is then obtained by integrating $\Gamma$ over the halo mass function:
\beq
\frac{d N_{\rm merge}}{dt dV} = \int d M_h \frac{d n_h}{d M_h} \Gamma(M_h).
\eeq
The halo mass function is well modeled on large scales $M_h \gtrsim 10^{10} M_{\odot}$, corresponding to $\sigma(M_h) \lesssim 4$ \cite{Tinker_08}. It is not unreasonable to extrapolate fitting functions several orders of magnitude in mass for standard adiabatic perturbations, as $\sigma(M_h)$ only depends logarithmically on $M_h$ in that case. However, for Poisson perturbations, $\sigma(M_h) \propto 1/\sqrt{M_h}$, and the mass function for very large values of $\sigma$ is anyone's guess. For lack of a better estimate we shall therefore simply use the Press-Schechter (PS) mass function \cite{PS}:
\barr
\frac{d n_h}{d M_h} &\approx& \sqrt{\frac2{\pi}} ~\frac{\overline{\rho}_{m}^0}{M_h} \Big{|}\frac{d \ln \sigma}{d M_h}\Big{|} \nu \rme^{- \frac12 \nu^2},  \ \ \nu \equiv \frac{1.68}{\sigma(M_h)},~
\earr
For the small halos of interest, with mass $M_h \ll 10^{12} M_{\odot}$, $\nu \ll 1$ and one can neglect the exponential term. We therefore arrive at
 \barr
 \frac{d N_{\rm merge}}{dt dV} \sim 600 \left(\overline{\rho}_m^0\right)^{73/42} \int_{M_c} d M_h \Big{|}\frac{d \sigma}{d M_h}\Big{|} \frac{\sigma^{3/14}}{M_h^{11/21}}, \ \ \label{eq:rate_bird}
 \earr
 where $M_c$ is a cutoff mass that truncates the otherwise divergent integral at $M_h \rightarrow 0$. To determine $M_c$, we follow BCM and require that the evaporation timescale of the halo is less than a Hubble time $t_0$. The evaporation time is approximately \cite{BT} 
\beq
t_{\rm evap} \sim 10 \frac{N}{\ln N} ~ t_h,
\eeq
where $t_h$ is the halo's crossing or dynamical time, and $N \equiv M/M_h$ is the number of objects in the halo. Using Eq.~\eqref{eq:rho_h}, we have 
\beq
t_h \sim 0.1 [\sigma(M_h)]^{-3/2} t_0.
\eeq
Therefore the cutoff halo mass $M_c$ is determined by the implicit equation 
\beq
[\sigma(M_c)]^{3/2} \sim \frac{M_c/M}{\ln (M_c/M)}. \label{eq:cutoff}
\eeq
Up to our specific assumptions about the the characteristic halo density and mass function, Eqs.~\eqref{eq:rate_bird} and \eqref{eq:cutoff} are fairly general. They illustrate that the merger rate is entirely determined by the variance $\sigma^2(M_h)$.

\subsection{Merger rate accounting for Poisson perturbations} 
 
Let us now explicitly include Poisson perturbations. From Eq.~\eqref{eq:sigma_M}, shot noise perturbations dominate over standard adiabatic perturbations for $M_h \leq M/\sigma_{\rm eq}^2 \approx 4 \times 10^4 M$. Provided this is the case, the variance at the present time is of order
\beq
\sigma(M_h) \approx z_{\rm eq} \sqrt{\frac{M}{M_h}}.
\eeq
Eq.~\eqref{eq:cutoff} therefore implies
\beq
z_{\rm eq}^{3/2} \left(\frac{M}{M_c}\right)^{3/4} \sim \frac{M_c/M}{\ln (M_c/M)},
\eeq
which implies
\beq
M_c \approx 3\times 10^3 ~ M \equiv N_c M,
\eeq
consistent with the results of Ref.~\citep{Afshordi_03}, and well into the regime where Poisson perturbations dominate over adiabatic ones. This is to be contrasted with the cutoff mass of BCM, who estimated $M_c \approx 400 M_{\odot}$ for $M = 30 M_{\odot}$, corresponding to $N_c \approx 13$.

The merger rate per halo, Eq.~\eqref{eq:Gamma_h}, is therefore
\barr
\Gamma &\sim&  3 \times 10^{10}  \left(\overline{\rho}_m^0\right)^{31/42} M_h^{10/21} (M/M_h)^{31/28} \nonumber\\
&\sim&  6 \times 10^{-14} \textrm{Gpc}^{-3} \textrm{yr}^{-1}  \left(\frac{m}{30}\right)^{10/21} \left(\frac{M_c}{M_h}\right)^{53/84}.~    \label{eq:rate_per_halo_2}
\earr
For $m = 30$, this rate is about two orders of magnitude larger than that of BCM at $M_h = M_c \approx 10^5 M_{\odot}$.

Finally, the total merger rate per unit volume is
\barr
\frac{d N_{\rm merge}}{dt dV} &\sim& 600 \left(\overline{\rho}_m^0\right)^{73/42} M^{-11/21} z_{\rm eq}^{17/14} \nonumber\\
&&\times \int_{N_c}^{\infty} \frac{dx}{x^{11/21}} \frac1{2 x^{3/2}} x^{-13/28} \nonumber\\
&\sim& 500 \left(\overline{\rho}_m^0\right)^{73/42} M^{-11/21} \nonumber\\
&\sim& 0.2 ~(m/30)^{-11/21} \textrm{Gpc}^{-3} \textrm{yr}^{-1}.
\earr
This is within a factor of a few of the result of BCM, as the much larger cutoff mass compensates for the enhanced merger rate per halo. 

We therefore conclude that $(i)$ the merger rate of PBH binaries formed in present-day halos is roughly consistent with the estimate of BCM, despite the important effect of small-scale Poisson fluctuations, and $(ii)$ it is a subdominant contribution to the overall PBH-binary merger rate\footnote{\yah{Ref.~\cite{Nishikawa:2017chy} recently argued that PBH mergers in dark matter spikes around supermassive black holes may yield an important contribution to the overall rate from present-day halos, possibly increasing the total rate by more than an order of magnitude. However, the associated uncertainties are very high and in any case this rate is still dwarfed by that of early-Universe binaries.}}. This last statement holds provided PBH binaries formed in the early Universe are not significantly disturbed between formation and merger, as our analytic estimates above indicate.

\section{Potential limits from existing LIGO observations} \label{sec:limits}

We now estimate upper limits on the volumetric merger rate of binary black holes set by LIGO O1, and how such limits would translate on the PBH abundance provided the merger rate is that computed in Section \ref{sec:initial_cond}.

In Ref.~\cite{LIGO_17}, the LIGO collaboration provides 90\% upper limits to the merger rate of intermediate-mass black holes, with individual masses up to 300 $M_{\odot}$. These limits depend on the spins of the black holes, in particular on their projection along the orbital angular momentum: in the case of $100-100 ~M_{\odot}$ binary, the upper bound varies by a factor $\sim 4$ between the nearly aligned and nearly anti-aligned cases. Since Ref.~\cite{LIGO_17} does not provide upper limits for non-zero spins for $M/M_{\odot} = 200$ and $300$, we shall use their zero-spin bounds for all cases, keeping in mind that they are only accurate up to a factor of a few.

For $M = 10, 20, 40~M_{\odot}$, we estimate the 90 \% upper limit on the merger rate from $R_{90\%} = - \ln(0.1)/\langle VT \rangle$ \cite{LIGO_17}, where $\langle VT \rangle$ is the average space-time volume to which the LIGO search is sensitive, and is obtained from integrating Fig.~7 of Ref.~\cite{LIGO_16bS}. We anticipate that LIGO also strongly constrains masses $M \leq 10~ M_{\odot}$, and defer this detailed analysis to the LIGO collaboration, updating that carried out in Ref.~\cite{LIGO_05} with the S2 run. We summarize our estimated limits in Table \ref{tab:limits}.

\begin{table}
\begin{tabular}{c|c}
  $M/M_{\odot}$ & $R_{90\%}$ [Gpc$^{-3}$ yr$^{-1}$]\\
  \hline\hline
  10 & 330\\
  20 & 77 \\
  40 & 15\\
  \hline
  100 & 2\\
  200 & 5\\
  300 & 20 \\
  \hline \hline
\end{tabular}
\caption{Estimated 90\% upper limits on the merger rate of equal-mass binary black holes from the LIGO O1 run. The limits for $M/M_{\odot} = 10, 20$ and $40$ are inferred from Refs.~\cite{LIGO_16b, LIGO_16bS}, and those for $M/M_{\odot} = 100, 200$ and $300$ are taken from Ref.~\cite{LIGO_17} for non-spinning black holes.}\label{tab:limits}
\end{table}

\begin{figure}
\includegraphics[width = \columnwidth]{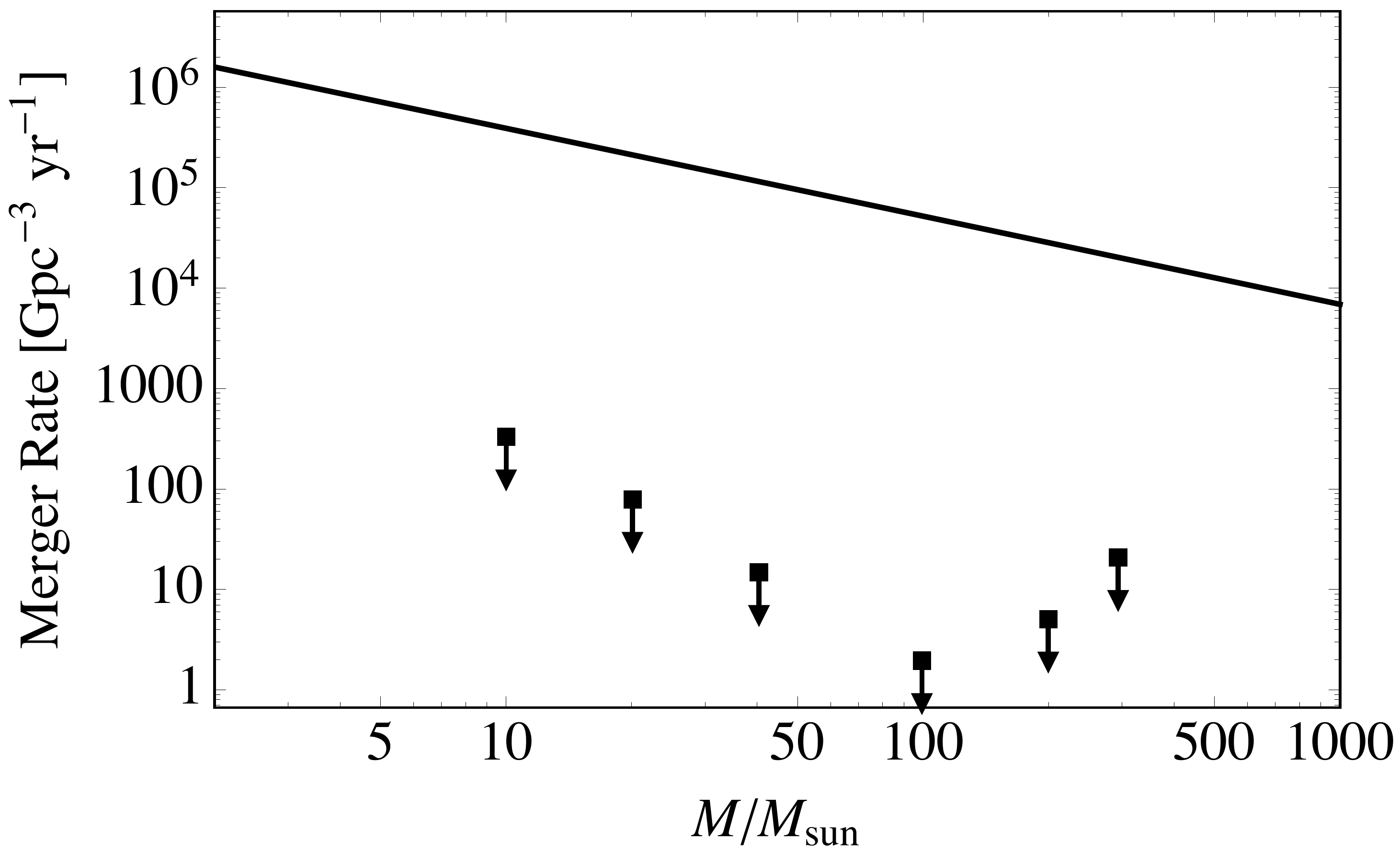}
\caption{Merger rate of PBH binaries if they make up all of the dark matter, and provided PBH binaries are not significantly perturbed between formation and merger (solid line). Superimposed are the upper limits from LIGO given in Table \ref{tab:limits} and described in the main text.} \label{fig:limits}
\end{figure}

We show these limits in Fig.~\ref{fig:limits}, alongside the PBH binary merger rate if they make all of the dark matter, and if PBH binaries are not significantly perturbed between formation and merger. We see that the latter \emph{largely} exceeds the estimated upper limits, by 3 to 4 orders of magnitude, depending on the mass. This indicates that LIGO could rule out PBHs as the dominant dark matter component, and set stringent upper limits to their abundance.

\begin{figure}
\includegraphics[width = \columnwidth]{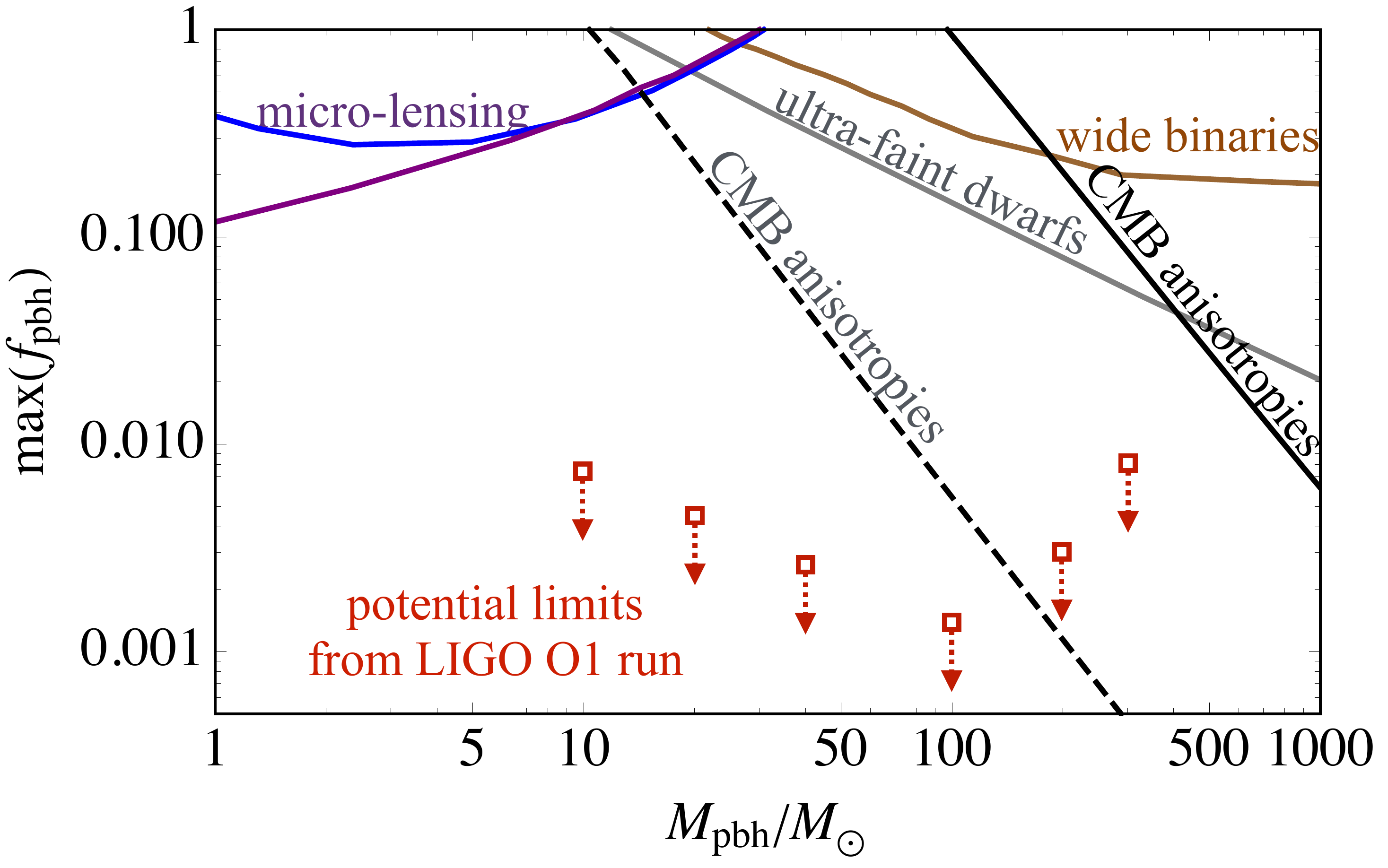}
\caption{Potential upper bounds on the fraction of dark matter in PBHs as a function of their mass, derived in this paper (red arrows), and assuming a narrow PBH mass function. These bounds need to be confirmed by numerical simulations. For comparison we also show the microlensing limits from the EROS \cite{Eros_07} (\emph{purple}) and MACHO \cite{Macho_01} (\emph{blue}) collaborations (see Ref.~\cite{Hawkins_15} for caveats and Ref.~\cite{Green_17} for a discussion of uncertainties), limits from wide Galactic binaries \cite{Monroy_14}, ultra-faint dwarf galaxies \cite{Brandt_16}, and CMB anisotropies \cite{YAH_17}.} \label{fig:bounds}
\end{figure}

To estimate these potential limits, we solve for the maximum PBH fraction for which the merger rate is below the LIGO upper limits. Note, that the merger rate is not linear in $f$, nor a simple power law through all range of $f$, so these limits must be computed numerically. We show the result in Fig.~\ref{fig:bounds}, alongside other existing bounds in that mass range. We see that LIGO O1 may limit PBHs to be no more than a percent of the dark matter for $M \sim 10-300~ M_{\odot}$. If confirmed with numerical computations, these would become the strongest existing bounds in that mass range.

\section{Discussion and conclusions} \label{sec:conclusions}

NSTT \cite{Nakamura_97} pointed out long ago that PBHs would form binaries in the early Universe, as a consequence of the chance proximity of PBH pairs, and estimated their merger rate at the present time. Following the first detection of a binary-black-hole merger \cite{LIGO_16}, Sasaki \emph{et al.} \cite{Sasaki_16} updated this calculation to $30~M_{\odot}$ PBHs, and generalized it to an arbitrary PBH abundance. They focused on the case where PBHs are a very subdominant fraction of the dark matter, as was implied by the stringent CMB spectral distortions bounds at the time \cite{Ricotti_08}, since then revised and significantly alleviated \cite{YAH_17} (see also \cite{Poulin_17}).

In this paper, we have, first of all, made several improvements to the calculation of NSST, and accurately computed the distribution of orbital parameters of PBH binaries forming in the early Universe. Specifically, we have computed the exact probability distribution of initial angular momentum for a close pair torqued by \emph{all} other PBHs, and have accounted for the tidal field of standard adiabatic density perturbations, dominant when PBHs make a small fraction of the dark matter. 

Our second and most important addition was to check thoroughly whether the highly eccentric orbits of PBH binaries merging today can get significantly disturbed between formation and merger. To do so, we have estimated the characteristic properties of the first non-linear structures, and as a consequence their effects on the orbital parameters of PBH binaries. We found that PBH binaries merging today are essentially unscathed by tidal torques and encounters with other PBHs. This robustness stems from the fact that these binaries typically form deep inside the radiation era and are very tight. We have also estimated the effect of baryon accretion to be much weaker than previous estimates \cite{Hayasaki_16}, but potentially important if unknown numerical prefactors happen to be large. 

Thirdly, we have revisited the calculation of Ref.~\cite{Bird_16} for the merger rate of PBH binaries forming in present-day halos through gravitational recombination. We have explicitly accounted for the previously neglected Poisson fluctuations resulting from the granularity of PBH dark matter. This shot noise greatly enhances the variance of density perturbations on small scales, and has pronounced effects on the properties of low-mass halos. We found that, despite a very different merger rate per halo and minimum halo mass, the final merger rate estimated by Ref.~\cite{Bird_16} was in the right ballpark. This makes it a very subdominant contribution to the overall binary-PBH merger rate.

Last, but not least, we have shown that the predicted merger rate would overwhelmingly exceed current upper bounds from LIGO O1 if PBHs make all of the dark matter. We have estimated possible upper bounds on the PBH abundance to be less than a percent of the dark matter in the range $10-300~M_{\odot}$, and as low as $\sim 0.2\%$ for $M = 100 ~M_{\odot}$. \caca{Let us point out that Ref.~\cite{Raidal_17} recently derived weaker potential upper limits on PBH abundance from LIGO. Our work differs in two aspects: first, we considered the rate of detectable merger events rather than the stochastic GW background as in Ref.~\cite{Raidal_17}. Second, and most importantly, we carefully studied the survival of PBH binaries between formation and the present time.}

These potentially stringent upper bounds need to be confirmed or refuted by a suite of numerical computations to check and complement our analytic estimates\footnote{This statement could and should be made for most published bounds on PBHs.}. Specifically, it would be interesting to investigate in more detail $(i)$ how the first PBH halos form and how they may affect PBH binaries, $(ii)$ the effects of a local bound halo of dark matter particles if PBHs are a subdominant dark matter component, and $(iii)$ baryon accretion onto a highy-eccentric PBH binary. These are complex but well-defined and interesting astrophysical problems, with possible implications beyond the topic of PBHs.

\section*{Acknowledgments}

We thank Scott Tremaine for discussions on tidal torquing of binaries and a careful reading of this manuscript. We are grateful to Andrew MacFadyen, Geoffrey Ryan and Kimitake Hayasaki for sharing thoughts and references on disk accretion physics. We also acknowledge conversations with Simeon Bird, Ilias Cholis, Juli\'{a}n Mu\~{n}oz, and Misao Sasaki. This work was supported at Johns Hopkins University by NSF Grant No. 0244990, NASA NNX15AB18G, and the Simons Foundation.

\appendix

\subsection*{Appendix: Probability distribution of the initial angular momentum} \label{app:j_dist}

\subsection{Torques by other PBHs} \label{app:j}

In this appendix we compute the full probability distribution of $\bs{j}$ resulting from torques by \emph{all} other PBHs, as given in Eq.~\eqref{eq:j-sum}. We apply Markoff's method, and parallel Chandrasekhar's derivation of the Holtsmark distribution for the gravitational field of point masses \cite{Chandra_43}. We consider $N$ masses uniformly distributed within a volume $V = \frac{4 \pi}{3} R^3$ and take the limit $N, V \rightarrow \infty$ at constant density $n = N/V$.

The vector $\bs{j}$ lies in the plane orthogonal to $\bs{x}$. Using Eq.~\eqref{eq:j-sum}, the two-dimensional probability distribution for $\bs{j}$ is formally given by
\barr
\frac{dP}{d^2j} = \underset{V \rightarrow \infty}{\lim} \prod_{p = 1}^N \int_V \frac{d^3 y_p}{V}   \delta_{\rm D} \left[\bs{j} -  1.5 \sum_{q=1}^N \frac{x^3}{y_q^5} y_{q ||}~ \bs{y}_{q \bot}\right],~~~~~~
\earr
where $y_{||} \equiv \bs{y} \cdot \hat{x}$ is the projection of $\bs{y}$ on $\hat{x}$, $\bs{y}_{\bot} \equiv \hat{x} \times \bs{y}$ is the (rotated) component of $\bs{y}$ perpendicular to $\hat{x}$, and $\delta_{\rm D}$ is the two-dimensional Dirac function, which we rewrite 
\beq
\delta_{\rm D}(\bs{X}) = \int_{\bs{k} \bot \hat{x}} \frac{d^2 k}{(2 \pi)^2} \rme^{i \bs{k} \cdot \bs{X}}.
\eeq
We hence get
\barr
\frac{dP}{d^2j} &=& \underset{V \rightarrow \infty}{\lim} \int \frac{d^2 k}{(2 \pi)^2}  \rme^{i \bs{k} \cdot \bs{j}} \mathcal{I}^N, \\
\mathcal{I} &\equiv& \int_V \frac{d^3 y}{V} \exp\left[- 1.5 i \frac{x^3}{y^5} y_{||} \bs{k} \cdot \bs{y}_{\bot}\right]\nonumber\\
 &=& 1 - \frac1{V} \int_V d^3 y \left\{1 - \exp\left[- 1.5 i \frac{x^3}{y^5} y_{||} \bs{k} \cdot \bs{y}_{\bot}\right]  \right\}.~~~~~~~
\earr
The latter integral is convergent when $V \rightarrow \infty$, so we get
\barr
\underset{V \rightarrow \infty}{\lim} \mathcal{I}^N &=& \underset{V \rightarrow \infty}{\lim} \left\{ 1 - \frac1{V} \int d^3 y \left[1 - \rme^{- 1.5 i \frac{x^3}{y^5} y_{||} \bs{k} \cdot \bs{y}_{\bot}} \right] \right\}^{n V} \nonumber\\
&=& \rme^{- n \mathcal{J}}, \\
\mathcal{J} &\equiv& \int d^3 y \left(1 - \exp\left[- 1.5 i \frac{x^3}{y^5} y_{||} \bs{k} \cdot \bs{y}_{\bot}\right] \right).
\earr
We rotate the component of $\bs{y}$ perpendicular to $\hat{x}$, $\bs{y} \rightarrow (\bs{y} \cdot \hat{x}) \hat{x} + \hat{x} \times \bs{y}$, and rescale $y \rightarrow (1.5 k)^{1/3} x y$, so the integral $\mathcal{J}$ becomes, recalling that $\bs{k} ~\bot~ \bs{x}$,
\barr
\mathcal{J} &=& 1.5 k x^3 \int d^3 y \left(1 - \exp\left[\frac{i}{y^3} (\hat{y} \cdot \hat{x}) (\hat{y} \cdot \hat{k})\right] \right) \nonumber\\
&=& 2 \pi k x^3 \int_0^{\infty} \frac{d v}{v^2}~ \frac{d^2 \hat{y}}{4 \pi} \left(1 - \rme^{i v (\hat{y} \cdot \hat{x}) (\hat{y} \cdot \hat{k})} \right),~~~~
\earr
where in the second line we changed variables to $v = 1/y^3$. Let us first consider the angular integral:
\barr
\mathcal{A}(v) &\equiv& \int \frac{d^2 \hat{y}}{4 \pi} \left(1- \rme^{i v (\hat{y} \cdot \hat{x}) (\hat{y} \cdot \hat{k})} \right) .
\earr
To compute it we use spherical polar coordinates, with polar axis $\hat{x} \times \hat{k}$:
\barr
\mathcal{A}(v) &=&  \int_0^{2 \pi} \frac{d \phi}{2\pi}\int_0^1 d \mu \left(1- \exp\left[\frac{i v}{2} \sin(2 \phi) (1- \mu^2) \right]\right)\nonumber\\
&=& \int_0^1 d \mu \left(1 - J_0\left[\frac{v}{2}(1-\mu^2)\right]\right),
\earr
where $J_0$ is the zeroth-order Bessel function. Since $J_0(x) = 1+ \mathcal{O}(x^2)$ for $x \rightarrow 0$, we can compute the integral over $v$ first:
\barr
\mathcal{J} &=& 2 \pi k x^3 \int_0^1 d \mu \int_0^{\infty} \frac{dv}{v^2} \left(1 - J_0\left[\frac{v}{2}(1-\mu^2)\right]\right) \nonumber\\
&=& 2 \pi k x^3  \int_0^1 d \mu \frac{1- \mu^2}{2} \int_0^{\infty} \frac{du}{u^2} (1 - J_0(u)).
\earr
The last two integrals are analytic, and we arrive at the simple expression
\beq
\mathcal{J} = \frac{2 \pi}{3} x^3 k = 0.5 \frac{X}{n} k. 
\eeq
We hence arrive at the probability distribution
\barr
\frac{dP}{dj} &=& 2 \pi j \frac{dP}{d^2 j} = j \int \frac{d^2 k}{2 \pi}  \rme^{i \bs{k} \cdot \bs{j} - j_X k}\nonumber\\
 &=&  j \int k dk J_0(k j) \rme^{- j_X k}, 
\earr
where $j_X \equiv 0.5 X$.
This integral is analytic and gives us Eqs.~\eqref{eq:dP_dj}-\eqref{eq:j_X}, which is the \emph{exact} expression for the distribution of reduced angular momentum (for a given $X$ hence $j_X$), accounting for tidal torquing by \emph{all} other PBHs (not just the nearest neighbor).

\subsection{Torques by density fluctuations} \label{sec:j2}

Let us now consider torques by linear density perturbations in the case where PBHs do not make all of the dark matter. The linear density field, hence tidal tensor, are Gaussian, and so is the resulting $\bs{j}$. Using Eq.~\eqref{eq:j-general}, the variance of $\bs{j}$ is given by,
\beq
\langle j^2 \rangle =  \frac{x^6}{4M^2} \left \langle \epsilon_{ijk} \hat{x}_j T_{kl} \hat{x}_l ~\epsilon_{ipq} \hat{x}_p T_{q m} \hat{x}_m  \right \rangle    
\eeq
where we have dropped the subscript ``eq" on $T_{ij}$. This expression simplifies to
\beq
\langle j^2 \rangle = \frac{x^6}{4M^2} \left \langle  T_{kl} \hat{x}_l T_{km} \hat{x}_m - (\hat{x}_k T_{kl} \hat{x}_l)^2  \right \rangle.
\eeq
Averaging over the direction of $\hat{x}$ we arrive at
\beq
\langle j^2 \rangle = \frac{x^6}{20 M^2} \left \langle  T_{ij} T_{ij} - \frac1{3} T_{ii} T_{jj}  \right \rangle.
\eeq
In Fourier space, $T_{ij} = - k_i k_j \phi = \hat{k}_i \hat{k}_j 4 \pi \overline{\rho}_{m, \rm eq} \delta_{\rm eq}$ , so we arrive at
\beq
\langle j^2 \rangle = \frac{3}{10} \left(\frac{4 \pi}3\right)^2 \frac{x^6}{M^2} [\overline{\rho}_{\rm eq}]^2 \langle \delta_{\rm eq}^2 \rangle,
\eeq
which implies Eq.~\eqref{eq:j-var}.

%
%
%

\bibliography{pbh_gw}

\begin{thebibliography}{78}%
\makeatletter
\providecommand \@ifxundefined [1]{%
 \@ifx{#1\undefined}
}%
\providecommand \@ifnum [1]{%
 \ifnum #1\expandafter \@firstoftwo
 \else \expandafter \@secondoftwo
 \fi
}%
\providecommand \@ifx [1]{%
 \ifx #1\expandafter \@firstoftwo
 \else \expandafter \@secondoftwo
 \fi
}%
\providecommand \natexlab [1]{#1}%
\providecommand \enquote  [1]{``#1''}%
\providecommand \bibnamefont  [1]{#1}%
\providecommand \bibfnamefont [1]{#1}%
\providecommand \citenamefont [1]{#1}%
\providecommand \href@noop [0]{\@secondoftwo}%
\providecommand \href [0]{\begingroup \@sanitize@url \@href}%
\providecommand \@href[1]{\@@startlink{#1}\@@href}%
\providecommand \@@href[1]{\endgroup#1\@@endlink}%
\providecommand \@sanitize@url [0]{\catcode `\\12\catcode `\$12\catcode
  `\&12\catcode `\#12\catcode `\^12\catcode `\_12\catcode `\%12\relax}%
\providecommand \@@startlink[1]{}%
\providecommand \@@endlink[0]{}%
\providecommand \url  [0]{\begingroup\@sanitize@url \@url }%
\providecommand \@url [1]{\endgroup\@href {#1}{\urlprefix }}%
\providecommand \urlprefix  [0]{URL }%
\providecommand \Eprint [0]{\href }%
\providecommand \doibase [0]{http://dx.doi.org/}%
\providecommand \selectlanguage [0]{\@gobble}%
\providecommand \bibinfo  [0]{\@secondoftwo}%
\providecommand \bibfield  [0]{\@secondoftwo}%
\providecommand \translation [1]{[#1]}%
\providecommand \BibitemOpen [0]{}%
\providecommand \bibitemStop [0]{}%
\providecommand \bibitemNoStop [0]{.\EOS\space}%
\providecommand \EOS [0]{\spacefactor3000\relax}%
\providecommand \BibitemShut  [1]{\csname bibitem#1\endcsname}%
\let\auto@bib@innerbib\@empty
\bibitem [{\citenamefont {{Marsh}}(2016)}]{Marsh_16}%
  \BibitemOpen
  \bibfield  {author} {\bibinfo {author} {\bibfnamefont {D.~J.~E.}\
  \bibnamefont {{Marsh}}},\ }\href {\doibase 10.1016/j.physrep.2016.06.005}
  {\bibfield  {journal} {\bibinfo  {journal} {\physrep}\ }\textbf {\bibinfo
  {volume} {643}},\ \bibinfo {pages} {1} (\bibinfo {year} {2016})},\ \Eprint
  {http://arxiv.org/abs/1510.07633} {arXiv:1510.07633} \BibitemShut {NoStop}%
\bibitem [{\citenamefont {{Jungman}}\ \emph {et~al.}(1996)\citenamefont
  {{Jungman}}, \citenamefont {{Kamionkowski}},\ and\ \citenamefont
  {{Griest}}}]{Jungman_96}%
  \BibitemOpen
  \bibfield  {author} {\bibinfo {author} {\bibfnamefont {G.}~\bibnamefont
  {{Jungman}}}, \bibinfo {author} {\bibfnamefont {M.}~\bibnamefont
  {{Kamionkowski}}}, \ and\ \bibinfo {author} {\bibfnamefont {K.}~\bibnamefont
  {{Griest}}},\ }\href {\doibase 10.1016/0370-1573(95)00058-5} {\bibfield
  {journal} {\bibinfo  {journal} {\physrep}\ }\textbf {\bibinfo {volume}
  {267}},\ \bibinfo {pages} {195} (\bibinfo {year} {1996})},\ \Eprint
  {http://arxiv.org/abs/hep-ph/9506380} {hep-ph/9506380} \BibitemShut {NoStop}%
\bibitem [{\citenamefont {{Griest}}(1991)}]{Griest_91}%
  \BibitemOpen
  \bibfield  {author} {\bibinfo {author} {\bibfnamefont {K.}~\bibnamefont
  {{Griest}}},\ }\href {\doibase 10.1086/169575} {\bibfield  {journal}
  {\bibinfo  {journal} {\apj}\ }\textbf {\bibinfo {volume} {366}},\ \bibinfo
  {pages} {412} (\bibinfo {year} {1991})}\BibitemShut {NoStop}%
\bibitem [{\citenamefont {Dolgov}\ \emph {et~al.}(2009)\citenamefont {Dolgov},
  \citenamefont {Kawasaki},\ and\ \citenamefont {Kevlishvili}}]{Dolgov:2008wu}%
  \BibitemOpen
  \bibfield  {author} {\bibinfo {author} {\bibfnamefont {A.~D.}\ \bibnamefont
  {Dolgov}}, \bibinfo {author} {\bibfnamefont {M.}~\bibnamefont {Kawasaki}}, \
  and\ \bibinfo {author} {\bibfnamefont {N.}~\bibnamefont {Kevlishvili}},\
  }\href {\doibase 10.1016/j.nuclphysb.2008.08.029} {\bibfield  {journal}
  {\bibinfo  {journal} {Nucl. Phys.}\ }\textbf {\bibinfo {volume} {B807}},\
  \bibinfo {pages} {229} (\bibinfo {year} {2009})},\ \Eprint
  {http://arxiv.org/abs/0806.2986} {arXiv:0806.2986 [hep-ph]} \BibitemShut
  {NoStop}%
\bibitem [{\citenamefont {Abbott}\ \emph
  {et~al.}(2016{\natexlab{a}})\citenamefont {Abbott} \emph {et~al.}}]{LIGO_16}%
  \BibitemOpen
  \bibfield  {author} {\bibinfo {author} {\bibfnamefont {B.~P.}\ \bibnamefont
  {Abbott}} \emph {et~al.} (\bibinfo {collaboration} {LIGO Scientific
  Collaboration and Virgo Collaboration}),\ }\href {\doibase
  10.1103/PhysRevLett.116.061102} {\bibfield  {journal} {\bibinfo  {journal}
  {Phys. Rev. Lett.}\ }\textbf {\bibinfo {volume} {116}},\ \bibinfo {pages}
  {061102} (\bibinfo {year} {2016}{\natexlab{a}})}\BibitemShut {NoStop}%
\bibitem [{\citenamefont {Abbott}\ \emph
  {et~al.}(2016{\natexlab{b}})\citenamefont {Abbott} \emph
  {et~al.}}]{GW151226}%
  \BibitemOpen
  \bibfield  {author} {\bibinfo {author} {\bibfnamefont {B.~P.}\ \bibnamefont
  {Abbott}} \emph {et~al.} (\bibinfo {collaboration} {LIGO Scientific
  Collaboration and Virgo Collaboration}),\ }\href {\doibase
  10.1103/PhysRevLett.116.241103} {\bibfield  {journal} {\bibinfo  {journal}
  {Phys. Rev. Lett.}\ }\textbf {\bibinfo {volume} {116}},\ \bibinfo {pages}
  {241103} (\bibinfo {year} {2016}{\natexlab{b}})}\BibitemShut {NoStop}%
\bibitem [{\citenamefont {Abbott}\ \emph
  {et~al.}(2017{\natexlab{a}})\citenamefont {Abbott} \emph
  {et~al.}}]{GW170104}%
  \BibitemOpen
  \bibfield  {author} {\bibinfo {author} {\bibfnamefont {B.~P.}\ \bibnamefont
  {Abbott}} \emph {et~al.} (\bibinfo {collaboration} {LIGO Scientific
  Collaboration and Virgo Collaboration}),\ }\href {\doibase
  10.1103/PhysRevLett.118.221101} {\bibfield  {journal} {\bibinfo  {journal}
  {Phys. Rev. Lett.}\ }\textbf {\bibinfo {volume} {118}},\ \bibinfo {pages}
  {221101} (\bibinfo {year} {2017}{\natexlab{a}})}\BibitemShut {NoStop}%
\bibitem [{\citenamefont {{Bird}}\ \emph {et~al.}(2016)\citenamefont {{Bird}},
  \citenamefont {{Cholis}}, \citenamefont {{Mu{\~n}oz}}, \citenamefont
  {{Ali-Ha{\"i}moud}}, \citenamefont {{Kamionkowski}}, \citenamefont
  {{Kovetz}}, \citenamefont {{Raccanelli}},\ and\ \citenamefont
  {{Riess}}}]{Bird_16}%
  \BibitemOpen
  \bibfield  {author} {\bibinfo {author} {\bibfnamefont {S.}~\bibnamefont
  {{Bird}}}, \bibinfo {author} {\bibfnamefont {I.}~\bibnamefont {{Cholis}}},
  \bibinfo {author} {\bibfnamefont {J.~B.}\ \bibnamefont {{Mu{\~n}oz}}},
  \bibinfo {author} {\bibfnamefont {Y.}~\bibnamefont {{Ali-Ha{\"i}moud}}},
  \bibinfo {author} {\bibfnamefont {M.}~\bibnamefont {{Kamionkowski}}},
  \bibinfo {author} {\bibfnamefont {E.~D.}\ \bibnamefont {{Kovetz}}}, \bibinfo
  {author} {\bibfnamefont {A.}~\bibnamefont {{Raccanelli}}}, \ and\ \bibinfo
  {author} {\bibfnamefont {A.~G.}\ \bibnamefont {{Riess}}},\ }\href {\doibase
  10.1103/PhysRevLett.116.201301} {\bibfield  {journal} {\bibinfo  {journal}
  {Phys. Rev. Lett.}\ }\textbf {\bibinfo {volume} {116}},\ \bibinfo {eid}
  {201301} (\bibinfo {year} {2016})},\ \Eprint
  {http://arxiv.org/abs/1603.00464} {arXiv:1603.00464} \BibitemShut {NoStop}%
\bibitem [{\citenamefont {{Sasaki}}\ \emph {et~al.}(2016)\citenamefont
  {{Sasaki}}, \citenamefont {{Suyama}}, \citenamefont {{Tanaka}},\ and\
  \citenamefont {{Yokoyama}}}]{Sasaki_16}%
  \BibitemOpen
  \bibfield  {author} {\bibinfo {author} {\bibfnamefont {M.}~\bibnamefont
  {{Sasaki}}}, \bibinfo {author} {\bibfnamefont {T.}~\bibnamefont {{Suyama}}},
  \bibinfo {author} {\bibfnamefont {T.}~\bibnamefont {{Tanaka}}}, \ and\
  \bibinfo {author} {\bibfnamefont {S.}~\bibnamefont {{Yokoyama}}},\ }\href
  {\doibase 10.1103/PhysRevLett.117.061101} {\bibfield  {journal} {\bibinfo
  {journal} {Phys. Rev. Lett.}\ }\textbf {\bibinfo {volume} {117}},\ \bibinfo
  {eid} {061101} (\bibinfo {year} {2016})},\ \Eprint
  {http://arxiv.org/abs/1603.08338} {arXiv:1603.08338} \BibitemShut {NoStop}%
\bibitem [{\citenamefont {Cholis}\ \emph {et~al.}(2016)\citenamefont {Cholis},
  \citenamefont {Kovetz}, \citenamefont {Ali-Ha{\"i}moud}, \citenamefont
  {Bird}, \citenamefont {Kamionkowski}, \citenamefont {Mu\~noz},\ and\
  \citenamefont {Raccanelli}}]{Cholis:2016kqi}%
  \BibitemOpen
  \bibfield  {author} {\bibinfo {author} {\bibfnamefont {I.}~\bibnamefont
  {Cholis}}, \bibinfo {author} {\bibfnamefont {E.~D.}\ \bibnamefont {Kovetz}},
  \bibinfo {author} {\bibfnamefont {Y.}~\bibnamefont {Ali-Ha{\"i}moud}},
  \bibinfo {author} {\bibfnamefont {S.}~\bibnamefont {Bird}}, \bibinfo {author}
  {\bibfnamefont {M.}~\bibnamefont {Kamionkowski}}, \bibinfo {author}
  {\bibfnamefont {J.~B.}\ \bibnamefont {Mu\~noz}}, \ and\ \bibinfo {author}
  {\bibfnamefont {A.}~\bibnamefont {Raccanelli}},\ }\href {\doibase
  10.1103/PhysRevD.94.084013} {\bibfield  {journal} {\bibinfo  {journal} {Phys.
  Rev. D}\ }\textbf {\bibinfo {volume} {94}},\ \bibinfo {pages} {084013}
  (\bibinfo {year} {2016})},\ \Eprint {http://arxiv.org/abs/1606.07437}
  {arXiv:1606.07437} \BibitemShut {NoStop}%
\bibitem [{\citenamefont {Clesse}\ and\ \citenamefont
  {Garc\'ia-Bellido}(2016)}]{Clesse:2016vqa}%
  \BibitemOpen
  \bibfield  {author} {\bibinfo {author} {\bibfnamefont {S.}~\bibnamefont
  {Clesse}}\ and\ \bibinfo {author} {\bibfnamefont {J.}~\bibnamefont
  {Garc\'ia-Bellido}},\ }\href {\doibase 10.1016/j.dark.2016.10.002} {\bibfield
   {journal} {\bibinfo  {journal} {Phys. Dark Univ.}\ }\textbf {\bibinfo
  {volume} {10}},\ \bibinfo {pages} {002} (\bibinfo {year} {2016})},\ \Eprint
  {http://arxiv.org/abs/1603.05234} {arXiv:1603.05234} \BibitemShut {NoStop}%
\bibitem [{\citenamefont {Raccanelli}\ \emph {et~al.}(2016)\citenamefont
  {Raccanelli}, \citenamefont {Kovetz}, \citenamefont {Bird}, \citenamefont
  {Cholis},\ and\ \citenamefont {Mu\~noz}}]{Raccanelli:2016cud}%
  \BibitemOpen
  \bibfield  {author} {\bibinfo {author} {\bibfnamefont {A.}~\bibnamefont
  {Raccanelli}}, \bibinfo {author} {\bibfnamefont {E.~D.}\ \bibnamefont
  {Kovetz}}, \bibinfo {author} {\bibfnamefont {S.}~\bibnamefont {Bird}},
  \bibinfo {author} {\bibfnamefont {I.}~\bibnamefont {Cholis}}, \ and\ \bibinfo
  {author} {\bibfnamefont {J.~B.}\ \bibnamefont {Mu\~noz}},\ }\href {\doibase
  10.1103/PhysRevD.94.023516} {\bibfield  {journal} {\bibinfo  {journal} {Phys.
  Rev.}\ }\textbf {\bibinfo {volume} {D94}},\ \bibinfo {pages} {023516}
  (\bibinfo {year} {2016})},\ \Eprint {http://arxiv.org/abs/1605.01405}
  {arXiv:1605.01405} \BibitemShut {NoStop}%
\bibitem [{\citenamefont {{Kovetz}}(2017)}]{Kovetz_17}%
  \BibitemOpen
  \bibfield  {author} {\bibinfo {author} {\bibfnamefont {E.~D.}\ \bibnamefont
  {{Kovetz}}},\ }\href@noop {} {\bibfield  {journal} {\bibinfo  {journal}
  {Phys.\ Rev.\ Lett., in press}\ } (\bibinfo {year} {2017})},\ \Eprint
  {http://arxiv.org/abs/1705.09182} {arXiv:1705.09182} \BibitemShut {NoStop}%
\bibitem [{\citenamefont {Blinnikov}\ \emph {et~al.}(2016)\citenamefont
  {Blinnikov}, \citenamefont {Dolgov}, \citenamefont {Porayko},\ and\
  \citenamefont {Postnov}}]{Blinnikov:2016bxu}%
  \BibitemOpen
  \bibfield  {author} {\bibinfo {author} {\bibfnamefont {S.}~\bibnamefont
  {Blinnikov}}, \bibinfo {author} {\bibfnamefont {A.}~\bibnamefont {Dolgov}},
  \bibinfo {author} {\bibfnamefont {N.~K.}\ \bibnamefont {Porayko}}, \ and\
  \bibinfo {author} {\bibfnamefont {K.}~\bibnamefont {Postnov}},\ }\href
  {\doibase 10.1088/1475-7516/2016/11/036} {\bibfield  {journal} {\bibinfo
  {journal} {JCAP}\ }\textbf {\bibinfo {volume} {1611}},\ \bibinfo {pages}
  {036} (\bibinfo {year} {2016})},\ \Eprint {http://arxiv.org/abs/1611.00541}
  {arXiv:1611.00541 [astro-ph.HE]} \BibitemShut {NoStop}%
\bibitem [{\citenamefont {Kashlinsky}(2016)}]{Kashlinsky:2016sdv}%
  \BibitemOpen
  \bibfield  {author} {\bibinfo {author} {\bibfnamefont {A.}~\bibnamefont
  {Kashlinsky}},\ }\href {\doibase 10.3847/2041-8205/823/2/L25} {\bibfield
  {journal} {\bibinfo  {journal} {Astrophys. J.}\ }\textbf {\bibinfo {volume}
  {823}},\ \bibinfo {pages} {L25} (\bibinfo {year} {2016})},\ \Eprint
  {http://arxiv.org/abs/1605.04023} {arXiv:1605.04023} \BibitemShut {NoStop}%
\bibitem [{\citenamefont {{Zel'dovich}}\ and\ \citenamefont
  {{Novikov}}(1967)}]{1967SvA....10..602Z}%
  \BibitemOpen
  \bibfield  {author} {\bibinfo {author} {\bibfnamefont {Y.~B.}\ \bibnamefont
  {{Zel'dovich}}}\ and\ \bibinfo {author} {\bibfnamefont {I.~D.}\ \bibnamefont
  {{Novikov}}},\ }\href@noop {} {\bibfield  {journal} {\bibinfo  {journal}
  {Soviet Astronomy}\ }\textbf {\bibinfo {volume} {10}},\ \bibinfo {pages}
  {602} (\bibinfo {year} {1967})}\BibitemShut {NoStop}%
\bibitem [{\citenamefont {{Hawking}}(1971)}]{Hawking_71}%
  \BibitemOpen
  \bibfield  {author} {\bibinfo {author} {\bibfnamefont {S.}~\bibnamefont
  {{Hawking}}},\ }\href {\doibase 10.1093/mnras/152.1.75} {\bibfield  {journal}
  {\bibinfo  {journal} {\mnras}\ }\textbf {\bibinfo {volume} {152}},\ \bibinfo
  {pages} {75} (\bibinfo {year} {1971})}\BibitemShut {NoStop}%
\bibitem [{\citenamefont {{Carr}}\ \emph {et~al.}(2010)\citenamefont {{Carr}},
  \citenamefont {{Kohri}}, \citenamefont {{Sendouda}},\ and\ \citenamefont
  {{Yokoyama}}}]{Carr_10}%
  \BibitemOpen
  \bibfield  {author} {\bibinfo {author} {\bibfnamefont {B.~J.}\ \bibnamefont
  {{Carr}}}, \bibinfo {author} {\bibfnamefont {K.}~\bibnamefont {{Kohri}}},
  \bibinfo {author} {\bibfnamefont {Y.}~\bibnamefont {{Sendouda}}}, \ and\
  \bibinfo {author} {\bibfnamefont {J.}~\bibnamefont {{Yokoyama}}},\ }\href
  {\doibase 10.1103/PhysRevD.81.104019} {\bibfield  {journal} {\bibinfo
  {journal} {\prd}\ }\textbf {\bibinfo {volume} {81}},\ \bibinfo {eid} {104019}
  (\bibinfo {year} {2010})},\ \Eprint {http://arxiv.org/abs/0912.5297}
  {arXiv:0912.5297} \BibitemShut {NoStop}%
\bibitem [{\citenamefont {{Carr}}\ \emph {et~al.}(2016)\citenamefont {{Carr}},
  \citenamefont {{K{\"u}hnel}},\ and\ \citenamefont {{Sandstad}}}]{Carr_16}%
  \BibitemOpen
  \bibfield  {author} {\bibinfo {author} {\bibfnamefont {B.}~\bibnamefont
  {{Carr}}}, \bibinfo {author} {\bibfnamefont {F.}~\bibnamefont
  {{K{\"u}hnel}}}, \ and\ \bibinfo {author} {\bibfnamefont {M.}~\bibnamefont
  {{Sandstad}}},\ }\href {\doibase 10.1103/PhysRevD.94.083504} {\bibfield
  {journal} {\bibinfo  {journal} {\prd}\ }\textbf {\bibinfo {volume} {94}},\
  \bibinfo {eid} {083504} (\bibinfo {year} {2016})},\ \Eprint
  {http://arxiv.org/abs/1607.06077} {arXiv:1607.06077} \BibitemShut {NoStop}%
\bibitem [{\citenamefont {{Alcock}}\ \emph {et~al.}(2001)\citenamefont
  {{Alcock}} \emph {et~al.}}]{Macho_01}%
  \BibitemOpen
  \bibfield  {author} {\bibinfo {author} {\bibfnamefont {C.}~\bibnamefont
  {{Alcock}}} \emph {et~al.},\ }\href {\doibase 10.1086/319636} {\bibfield
  {journal} {\bibinfo  {journal} {\apjl}\ }\textbf {\bibinfo {volume} {550}},\
  \bibinfo {pages} {L169} (\bibinfo {year} {2001})},\ \Eprint
  {http://arxiv.org/abs/astro-ph/0011506} {astro-ph/0011506} \BibitemShut
  {NoStop}%
\bibitem [{\citenamefont {{Tisserand}}\ \emph {et~al.}(2007)\citenamefont
  {{Tisserand}} \emph {et~al.}}]{Eros_07}%
  \BibitemOpen
  \bibfield  {author} {\bibinfo {author} {\bibfnamefont {P.}~\bibnamefont
  {{Tisserand}}} \emph {et~al.},\ }\href {\doibase 10.1051/0004-6361:20066017}
  {\bibfield  {journal} {\bibinfo  {journal} {\aap}\ }\textbf {\bibinfo
  {volume} {469}},\ \bibinfo {pages} {387} (\bibinfo {year} {2007})},\ \Eprint
  {http://arxiv.org/abs/astro-ph/0607207} {astro-ph/0607207} \BibitemShut
  {NoStop}%
\bibitem [{\citenamefont {{Monroy-Rodr{\'{\i}}guez}}\ and\ \citenamefont
  {{Allen}}(2014)}]{Monroy_14}%
  \BibitemOpen
  \bibfield  {author} {\bibinfo {author} {\bibfnamefont {M.~A.}\ \bibnamefont
  {{Monroy-Rodr{\'{\i}}guez}}}\ and\ \bibinfo {author} {\bibfnamefont
  {C.}~\bibnamefont {{Allen}}},\ }\href {\doibase 10.1088/0004-637X/790/2/159}
  {\bibfield  {journal} {\bibinfo  {journal} {\apj}\ }\textbf {\bibinfo
  {volume} {790}},\ \bibinfo {eid} {159} (\bibinfo {year} {2014})},\ \Eprint
  {http://arxiv.org/abs/1406.5169} {arXiv:1406.5169} \BibitemShut {NoStop}%
\bibitem [{\citenamefont {{Ricotti}}\ \emph {et~al.}(2008)\citenamefont
  {{Ricotti}}, \citenamefont {{Ostriker}},\ and\ \citenamefont
  {{Mack}}}]{Ricotti_08}%
  \BibitemOpen
  \bibfield  {author} {\bibinfo {author} {\bibfnamefont {M.}~\bibnamefont
  {{Ricotti}}}, \bibinfo {author} {\bibfnamefont {J.~P.}\ \bibnamefont
  {{Ostriker}}}, \ and\ \bibinfo {author} {\bibfnamefont {K.~J.}\ \bibnamefont
  {{Mack}}},\ }\href {\doibase 10.1086/587831} {\bibfield  {journal} {\bibinfo
  {journal} {\apj}\ }\textbf {\bibinfo {volume} {680}},\ \bibinfo {eid}
  {829-845} (\bibinfo {year} {2008})},\ \Eprint
  {http://arxiv.org/abs/0709.0524} {arXiv:0709.0524} \BibitemShut {NoStop}%
\bibitem [{\citenamefont {{Ali-Ha{\"i}moud}}\ and\ \citenamefont
  {{Kamionkowski}}(2017)}]{YAH_17}%
  \BibitemOpen
  \bibfield  {author} {\bibinfo {author} {\bibfnamefont {Y.}~\bibnamefont
  {{Ali-Ha{\"i}moud}}}\ and\ \bibinfo {author} {\bibfnamefont {M.}~\bibnamefont
  {{Kamionkowski}}},\ }\href {\doibase 10.1103/PhysRevD.95.043534} {\bibfield
  {journal} {\bibinfo  {journal} {\prd}\ }\textbf {\bibinfo {volume} {95}},\
  \bibinfo {eid} {043534} (\bibinfo {year} {2017})},\ \Eprint
  {http://arxiv.org/abs/1612.05644} {arXiv:1612.05644} \BibitemShut {NoStop}%
\bibitem [{\citenamefont {{Brandt}}(2016)}]{Brandt_16}%
  \BibitemOpen
  \bibfield  {author} {\bibinfo {author} {\bibfnamefont {T.~D.}\ \bibnamefont
  {{Brandt}}},\ }\href {\doibase 10.3847/2041-8205/824/2/L31} {\bibfield
  {journal} {\bibinfo  {journal} {\apjl}\ }\textbf {\bibinfo {volume} {824}},\
  \bibinfo {eid} {L31} (\bibinfo {year} {2016})},\ \Eprint
  {http://arxiv.org/abs/1605.03665} {arXiv:1605.03665} \BibitemShut {NoStop}%
\bibitem [{\citenamefont {{Koushiappas}}\ and\ \citenamefont
  {{Loeb}}(2017)}]{Koushiappas_17}%
  \BibitemOpen
  \bibfield  {author} {\bibinfo {author} {\bibfnamefont {S.~M.}\ \bibnamefont
  {{Koushiappas}}}\ and\ \bibinfo {author} {\bibfnamefont {A.}~\bibnamefont
  {{Loeb}}},\ }\href@noop {} {\bibfield  {journal} {\bibinfo  {journal} {ArXiv
  e-prints}\ } (\bibinfo {year} {2017})},\ \Eprint
  {http://arxiv.org/abs/1704.01668} {arXiv:1704.01668} \BibitemShut {NoStop}%
\bibitem [{\citenamefont {Li}\ \emph {et~al.}(2017)\citenamefont {Li} \emph
  {et~al.}}]{Li:2016utv}%
  \BibitemOpen
  \bibfield  {author} {\bibinfo {author} {\bibfnamefont {T.~S.}\ \bibnamefont
  {Li}} \emph {et~al.} (\bibinfo {collaboration} {DES}),\ }\href {\doibase
  10.3847/1538-4357/aa6113} {\bibfield  {journal} {\bibinfo  {journal}
  {Astrophys. J.}\ }\textbf {\bibinfo {volume} {838}},\ \bibinfo {pages} {8}
  (\bibinfo {year} {2017})},\ \Eprint {http://arxiv.org/abs/1611.05052}
  {arXiv:1611.05052 [astro-ph.GA]} \BibitemShut {NoStop}%
\bibitem [{\citenamefont {Gaggero}\ \emph {et~al.}(2017)\citenamefont
  {Gaggero}, \citenamefont {Bertone}, \citenamefont {Calore}, \citenamefont
  {Connors}, \citenamefont {Lovell}, \citenamefont {Markoff},\ and\
  \citenamefont {Storm}}]{Gaggero:2016dpq}%
  \BibitemOpen
  \bibfield  {author} {\bibinfo {author} {\bibfnamefont {D.}~\bibnamefont
  {Gaggero}}, \bibinfo {author} {\bibfnamefont {G.}~\bibnamefont {Bertone}},
  \bibinfo {author} {\bibfnamefont {F.}~\bibnamefont {Calore}}, \bibinfo
  {author} {\bibfnamefont {R.~M.~T.}\ \bibnamefont {Connors}}, \bibinfo
  {author} {\bibfnamefont {M.}~\bibnamefont {Lovell}}, \bibinfo {author}
  {\bibfnamefont {S.}~\bibnamefont {Markoff}}, \ and\ \bibinfo {author}
  {\bibfnamefont {E.}~\bibnamefont {Storm}},\ }\href {\doibase
  10.1103/PhysRevLett.118.241101} {\bibfield  {journal} {\bibinfo  {journal}
  {Phys. Rev. Lett.}\ }\textbf {\bibinfo {volume} {118}},\ \bibinfo {pages}
  {241101} (\bibinfo {year} {2017})},\ \Eprint
  {http://arxiv.org/abs/1612.00457} {arXiv:1612.00457 [astro-ph.HE]}
  \BibitemShut {NoStop}%
\bibitem [{\citenamefont {Inoue}\ and\ \citenamefont
  {Kusenko}(2017)}]{Inoue:2017csr}%
  \BibitemOpen
  \bibfield  {author} {\bibinfo {author} {\bibfnamefont {Y.}~\bibnamefont
  {Inoue}}\ and\ \bibinfo {author} {\bibfnamefont {A.}~\bibnamefont
  {Kusenko}},\ }\href@noop {} {\  (\bibinfo {year} {2017})},\ \Eprint
  {http://arxiv.org/abs/1705.00791} {arXiv:1705.00791 [astro-ph.CO]}
  \BibitemShut {NoStop}%
\bibitem [{\citenamefont {Agol}\ and\ \citenamefont
  {Kamionkowski}(2002)}]{Agol:2001hb}%
  \BibitemOpen
  \bibfield  {author} {\bibinfo {author} {\bibfnamefont {E.}~\bibnamefont
  {Agol}}\ and\ \bibinfo {author} {\bibfnamefont {M.}~\bibnamefont
  {Kamionkowski}},\ }\href {\doibase 10.1046/j.1365-8711.2002.05523.x}
  {\bibfield  {journal} {\bibinfo  {journal} {Mon. Not. Roy. Astron. Soc.}\
  }\textbf {\bibinfo {volume} {334}},\ \bibinfo {pages} {553} (\bibinfo {year}
  {2002})},\ \Eprint {http://arxiv.org/abs/astro-ph/0109539}
  {arXiv:astro-ph/0109539 [astro-ph]} \BibitemShut {NoStop}%
\bibitem [{\citenamefont {{Carr}}\ \emph {et~al.}(2017)\citenamefont {{Carr}},
  \citenamefont {{Raidal}}, \citenamefont {{Tenkanen}}, \citenamefont
  {{Vaskonen}},\ and\ \citenamefont {{Veerm{\"a}e}}}]{Carr_17}%
  \BibitemOpen
  \bibfield  {author} {\bibinfo {author} {\bibfnamefont {B.}~\bibnamefont
  {{Carr}}}, \bibinfo {author} {\bibfnamefont {M.}~\bibnamefont {{Raidal}}},
  \bibinfo {author} {\bibfnamefont {T.}~\bibnamefont {{Tenkanen}}}, \bibinfo
  {author} {\bibfnamefont {V.}~\bibnamefont {{Vaskonen}}}, \ and\ \bibinfo
  {author} {\bibfnamefont {H.}~\bibnamefont {{Veerm{\"a}e}}},\ }\href@noop {}
  {\bibfield  {journal} {\bibinfo  {journal} {ArXiv e-prints}\ } (\bibinfo
  {year} {2017})},\ \Eprint {http://arxiv.org/abs/1705.05567}
  {arXiv:1705.05567} \BibitemShut {NoStop}%
\bibitem [{\citenamefont {{Green}}(2017)}]{Green_17}%
  \BibitemOpen
  \bibfield  {author} {\bibinfo {author} {\bibfnamefont {A.~M.}\ \bibnamefont
  {{Green}}},\ }\href@noop {} {\bibfield  {journal} {\bibinfo  {journal} {ArXiv
  e-prints}\ } (\bibinfo {year} {2017})},\ \Eprint
  {http://arxiv.org/abs/1705.10818} {arXiv:1705.10818} \BibitemShut {NoStop}%
\bibitem [{\citenamefont {{Poulin}}\ \emph {et~al.}(2017)\citenamefont
  {{Poulin}}, \citenamefont {{Serpico}}, \citenamefont {{Calore}},
  \citenamefont {{Clesse}},\ and\ \citenamefont {{Kohri}}}]{Poulin_17}%
  \BibitemOpen
  \bibfield  {author} {\bibinfo {author} {\bibfnamefont {V.}~\bibnamefont
  {{Poulin}}}, \bibinfo {author} {\bibfnamefont {P.~D.}\ \bibnamefont
  {{Serpico}}}, \bibinfo {author} {\bibfnamefont {F.}~\bibnamefont {{Calore}}},
  \bibinfo {author} {\bibfnamefont {S.}~\bibnamefont {{Clesse}}}, \ and\
  \bibinfo {author} {\bibfnamefont {K.}~\bibnamefont {{Kohri}}},\ }\href@noop
  {} {\bibfield  {journal} {\bibinfo  {journal} {ArXiv e-prints}\ } (\bibinfo
  {year} {2017})},\ \Eprint {http://arxiv.org/abs/1707.04206}
  {arXiv:1707.04206} \BibitemShut {NoStop}%
\bibitem [{\citenamefont {{Garc{\'{\i}}a-Bellido}}(2017)}]{Garcia-Bellido_17}%
  \BibitemOpen
  \bibfield  {author} {\bibinfo {author} {\bibfnamefont {J.}~\bibnamefont
  {{Garc{\'{\i}}a-Bellido}}},\ }\href@noop {} {\bibfield  {journal} {\bibinfo
  {journal} {ArXiv e-prints}\ } (\bibinfo {year} {2017})},\ \Eprint
  {http://arxiv.org/abs/1702.08275} {arXiv:1702.08275} \BibitemShut {NoStop}%
\bibitem [{\citenamefont {{Kovetz}}\ \emph {et~al.}(2017)\citenamefont
  {{Kovetz}}, \citenamefont {{Cholis}}, \citenamefont {{Breysse}},\ and\
  \citenamefont {{Kamionkowski}}}]{Kovetz:2016kpi}%
  \BibitemOpen
  \bibfield  {author} {\bibinfo {author} {\bibfnamefont {E.~D.}\ \bibnamefont
  {{Kovetz}}}, \bibinfo {author} {\bibfnamefont {I.}~\bibnamefont {{Cholis}}},
  \bibinfo {author} {\bibfnamefont {P.~C.}\ \bibnamefont {{Breysse}}}, \ and\
  \bibinfo {author} {\bibfnamefont {M.}~\bibnamefont {{Kamionkowski}}},\ }\href
  {\doibase 10.1103/PhysRevD.95.103010} {\bibfield  {journal} {\bibinfo
  {journal} {\prd}\ }\textbf {\bibinfo {volume} {95}},\ \bibinfo {eid} {103010}
  (\bibinfo {year} {2017})},\ \Eprint {http://arxiv.org/abs/1611.01157}
  {arXiv:1611.01157} \BibitemShut {NoStop}%
\bibitem [{\citenamefont {Mu\~noz}\ \emph {et~al.}(2016)\citenamefont
  {Mu\~noz}, \citenamefont {Kovetz}, \citenamefont {Dai},\ and\ \citenamefont
  {Kamionkowski}}]{Munoz:2016tmg}%
  \BibitemOpen
  \bibfield  {author} {\bibinfo {author} {\bibfnamefont {J.~B.}\ \bibnamefont
  {Mu\~noz}}, \bibinfo {author} {\bibfnamefont {E.~D.}\ \bibnamefont {Kovetz}},
  \bibinfo {author} {\bibfnamefont {L.}~\bibnamefont {Dai}}, \ and\ \bibinfo
  {author} {\bibfnamefont {M.}~\bibnamefont {Kamionkowski}},\ }\href {\doibase
  10.1103/PhysRevLett.117.091301} {\bibfield  {journal} {\bibinfo  {journal}
  {Phys. Rev. Lett.}\ }\textbf {\bibinfo {volume} {117}},\ \bibinfo {pages}
  {091301} (\bibinfo {year} {2016})},\ \Eprint
  {http://arxiv.org/abs/1605.00008} {arXiv:1605.00008} \BibitemShut {NoStop}%
\bibitem [{\citenamefont {Schutz}\ and\ \citenamefont
  {Liu}(2017)}]{Schutz:2016khr}%
  \BibitemOpen
  \bibfield  {author} {\bibinfo {author} {\bibfnamefont {K.}~\bibnamefont
  {Schutz}}\ and\ \bibinfo {author} {\bibfnamefont {A.}~\bibnamefont {Liu}},\
  }\href {\doibase 10.1103/PhysRevD.95.023002} {\bibfield  {journal} {\bibinfo
  {journal} {Phys. Rev.}\ }\textbf {\bibinfo {volume} {D95}},\ \bibinfo {pages}
  {023002} (\bibinfo {year} {2017})},\ \Eprint
  {http://arxiv.org/abs/1610.04234} {arXiv:1610.04234 [astro-ph.CO]}
  \BibitemShut {NoStop}%
\bibitem [{\citenamefont {{Raidal}}\ \emph {et~al.}(2017)\citenamefont
  {{Raidal}}, \citenamefont {{Vaskonen}},\ and\ \citenamefont
  {{Veerm{\"a}e}}}]{Raidal_17}%
  \BibitemOpen
  \bibfield  {author} {\bibinfo {author} {\bibfnamefont {M.}~\bibnamefont
  {{Raidal}}}, \bibinfo {author} {\bibfnamefont {V.}~\bibnamefont
  {{Vaskonen}}}, \ and\ \bibinfo {author} {\bibfnamefont {H.}~\bibnamefont
  {{Veerm{\"a}e}}},\ }\href@noop {} {\bibfield  {journal} {\bibinfo  {journal}
  {ArXiv e-prints}\ } (\bibinfo {year} {2017})},\ \Eprint
  {http://arxiv.org/abs/1707.01480} {arXiv:1707.01480} \BibitemShut {NoStop}%
\bibitem [{\citenamefont {{Nakamura}}\ \emph {et~al.}(1997)\citenamefont
  {{Nakamura}}, \citenamefont {{Sasaki}}, \citenamefont {{Tanaka}},\ and\
  \citenamefont {{Thorne}}}]{Nakamura_97}%
  \BibitemOpen
  \bibfield  {author} {\bibinfo {author} {\bibfnamefont {T.}~\bibnamefont
  {{Nakamura}}}, \bibinfo {author} {\bibfnamefont {M.}~\bibnamefont
  {{Sasaki}}}, \bibinfo {author} {\bibfnamefont {T.}~\bibnamefont {{Tanaka}}},
  \ and\ \bibinfo {author} {\bibfnamefont {K.~S.}\ \bibnamefont {{Thorne}}},\
  }\href {\doibase 10.1086/310886} {\bibfield  {journal} {\bibinfo  {journal}
  {\apjl}\ }\textbf {\bibinfo {volume} {487}},\ \bibinfo {pages} {L139}
  (\bibinfo {year} {1997})},\ \Eprint {http://arxiv.org/abs/astro-ph/9708060}
  {astro-ph/9708060} \BibitemShut {NoStop}%
\bibitem [{\citenamefont {Abbott}\ \emph {et~al.}(2005)\citenamefont {Abbott}
  \emph {et~al.}}]{LIGO_05}%
  \BibitemOpen
  \bibfield  {author} {\bibinfo {author} {\bibfnamefont {B.~P.}\ \bibnamefont
  {Abbott}} \emph {et~al.} (\bibinfo {collaboration} {LIGO Scientific
  Collaboration}),\ }\href {\doibase 10.1103/PhysRevD.72.082002} {\bibfield
  {journal} {\bibinfo  {journal} {\prd}\ }\textbf {\bibinfo {volume} {72}},\
  \bibinfo {eid} {082002} (\bibinfo {year} {2005})},\ \Eprint
  {http://arxiv.org/abs/gr-qc/0505042} {gr-qc/0505042} \BibitemShut {NoStop}%
\bibitem [{\citenamefont {{Ioka}}\ \emph {et~al.}(1998)\citenamefont {{Ioka}},
  \citenamefont {{Chiba}}, \citenamefont {{Tanaka}},\ and\ \citenamefont
  {{Nakamura}}}]{Ioka_98}%
  \BibitemOpen
  \bibfield  {author} {\bibinfo {author} {\bibfnamefont {K.}~\bibnamefont
  {{Ioka}}}, \bibinfo {author} {\bibfnamefont {T.}~\bibnamefont {{Chiba}}},
  \bibinfo {author} {\bibfnamefont {T.}~\bibnamefont {{Tanaka}}}, \ and\
  \bibinfo {author} {\bibfnamefont {T.}~\bibnamefont {{Nakamura}}},\ }\href
  {\doibase 10.1103/PhysRevD.58.063003} {\bibfield  {journal} {\bibinfo
  {journal} {\prd}\ }\textbf {\bibinfo {volume} {58}},\ \bibinfo {eid} {063003}
  (\bibinfo {year} {1998})},\ \Eprint {http://arxiv.org/abs/astro-ph/9807018}
  {astro-ph/9807018} \BibitemShut {NoStop}%
\bibitem [{\citenamefont {Ade}\ \emph {et~al.}(2016)\citenamefont {Ade} \emph
  {et~al.}}]{Planck_16}%
  \BibitemOpen
  \bibfield  {author} {\bibinfo {author} {\bibfnamefont {P.~A.~R.}\
  \bibnamefont {Ade}} \emph {et~al.} (\bibinfo {collaboration} {Planck
  Collaboration}),\ }\href {\doibase 10.1051/0004-6361/201525830} {\bibfield
  {journal} {\bibinfo  {journal} {\aap}\ }\textbf {\bibinfo {volume} {594}},\
  \bibinfo {eid} {A13} (\bibinfo {year} {2016})},\ \Eprint
  {http://arxiv.org/abs/1502.01589} {arXiv:1502.01589} \BibitemShut {NoStop}%
\bibitem [{\citenamefont {{Chisholm}}(2006)}]{Chisholm_06}%
  \BibitemOpen
  \bibfield  {author} {\bibinfo {author} {\bibfnamefont {J.~R.}\ \bibnamefont
  {{Chisholm}}},\ }\href {\doibase 10.1103/PhysRevD.73.083504} {\bibfield
  {journal} {\bibinfo  {journal} {\prd}\ }\textbf {\bibinfo {volume} {73}},\
  \bibinfo {eid} {083504} (\bibinfo {year} {2006})},\ \Eprint
  {http://arxiv.org/abs/astro-ph/0509141} {astro-ph/0509141} \BibitemShut
  {NoStop}%
\bibitem [{\citenamefont {{Eroshenko}}(2016{\natexlab{a}})}]{Eroshenko_16}%
  \BibitemOpen
  \bibfield  {author} {\bibinfo {author} {\bibfnamefont {Y.~N.}\ \bibnamefont
  {{Eroshenko}}},\ }\href@noop {} {\bibfield  {journal} {\bibinfo  {journal}
  {ArXiv e-prints}\ } (\bibinfo {year} {2016}{\natexlab{a}})},\ \Eprint
  {http://arxiv.org/abs/1604.04932} {arXiv:1604.04932} \BibitemShut {NoStop}%
\bibitem [{\citenamefont {{Hayasaki}}\ \emph {et~al.}(2016)\citenamefont
  {{Hayasaki}}, \citenamefont {{Takahashi}}, \citenamefont {{Sendouda}},\ and\
  \citenamefont {{Nagataki}}}]{Hayasaki_16}%
  \BibitemOpen
  \bibfield  {author} {\bibinfo {author} {\bibfnamefont {K.}~\bibnamefont
  {{Hayasaki}}}, \bibinfo {author} {\bibfnamefont {K.}~\bibnamefont
  {{Takahashi}}}, \bibinfo {author} {\bibfnamefont {Y.}~\bibnamefont
  {{Sendouda}}}, \ and\ \bibinfo {author} {\bibfnamefont {S.}~\bibnamefont
  {{Nagataki}}},\ }\href {\doibase 10.1093/pasj/psw065} {\bibfield  {journal}
  {\bibinfo  {journal} {\pasj}\ }\textbf {\bibinfo {volume} {68}},\ \bibinfo
  {eid} {66} (\bibinfo {year} {2016})},\ \Eprint
  {http://arxiv.org/abs/0909.1738} {arXiv:0909.1738} \BibitemShut {NoStop}%
\bibitem [{\citenamefont {{Peters}}(1964)}]{Peters_64}%
  \BibitemOpen
  \bibfield  {author} {\bibinfo {author} {\bibfnamefont {P.~C.}\ \bibnamefont
  {{Peters}}},\ }\href {\doibase 10.1103/PhysRev.136.B1224} {\bibfield
  {journal} {\bibinfo  {journal} {Physical Review}\ }\textbf {\bibinfo {volume}
  {136}},\ \bibinfo {pages} {1224} (\bibinfo {year} {1964})}\BibitemShut
  {NoStop}%
\bibitem [{\citenamefont {{Ir{\v s}i{\v c}}}\ \emph {et~al.}(2017)\citenamefont
  {{Ir{\v s}i{\v c}}}, \citenamefont {{Viel}}, \citenamefont {{Haehnelt}},
  \citenamefont {{Bolton}}, \citenamefont {{Cristiani}}, \citenamefont
  {{Becker}}, \citenamefont {{D'Odorico}}, \citenamefont {{Cupani}},
  \citenamefont {{Kim}}, \citenamefont {{Berg}}, \citenamefont {{L{\'o}pez}},
  \citenamefont {{Ellison}}, \citenamefont {{Christensen}}, \citenamefont
  {{Denney}},\ and\ \citenamefont {{Worseck}}}]{Irsic_17}%
  \BibitemOpen
  \bibfield  {author} {\bibinfo {author} {\bibfnamefont {V.}~\bibnamefont
  {{Ir{\v s}i{\v c}}}}, \bibinfo {author} {\bibfnamefont {M.}~\bibnamefont
  {{Viel}}}, \bibinfo {author} {\bibfnamefont {M.~G.}\ \bibnamefont
  {{Haehnelt}}}, \bibinfo {author} {\bibfnamefont {J.~S.}\ \bibnamefont
  {{Bolton}}}, \bibinfo {author} {\bibfnamefont {S.}~\bibnamefont
  {{Cristiani}}}, \bibinfo {author} {\bibfnamefont {G.~D.}\ \bibnamefont
  {{Becker}}}, \bibinfo {author} {\bibfnamefont {V.}~\bibnamefont
  {{D'Odorico}}}, \bibinfo {author} {\bibfnamefont {G.}~\bibnamefont
  {{Cupani}}}, \bibinfo {author} {\bibfnamefont {T.-S.}\ \bibnamefont {{Kim}}},
  \bibinfo {author} {\bibfnamefont {T.~A.~M.}\ \bibnamefont {{Berg}}}, \bibinfo
  {author} {\bibfnamefont {S.}~\bibnamefont {{L{\'o}pez}}}, \bibinfo {author}
  {\bibfnamefont {S.}~\bibnamefont {{Ellison}}}, \bibinfo {author}
  {\bibfnamefont {L.}~\bibnamefont {{Christensen}}}, \bibinfo {author}
  {\bibfnamefont {K.~D.}\ \bibnamefont {{Denney}}}, \ and\ \bibinfo {author}
  {\bibfnamefont {G.}~\bibnamefont {{Worseck}}},\ }\href {\doibase
  10.1103/PhysRevD.96.023522} {\bibfield  {journal} {\bibinfo  {journal}
  {\prd}\ }\textbf {\bibinfo {volume} {96}},\ \bibinfo {eid} {023522} (\bibinfo
  {year} {2017})},\ \Eprint {http://arxiv.org/abs/1702.01764}
  {arXiv:1702.01764} \BibitemShut {NoStop}%
\bibitem [{\citenamefont {{Misner}}\ \emph {et~al.}(1973)\citenamefont
  {{Misner}}, \citenamefont {{Thorne}},\ and\ \citenamefont {{Wheeler}}}]{MTW}%
  \BibitemOpen
  \bibfield  {author} {\bibinfo {author} {\bibfnamefont {C.~W.}\ \bibnamefont
  {{Misner}}}, \bibinfo {author} {\bibfnamefont {K.~S.}\ \bibnamefont
  {{Thorne}}}, \ and\ \bibinfo {author} {\bibfnamefont {J.~A.}\ \bibnamefont
  {{Wheeler}}},\ }\href@noop {} {\emph {\bibinfo {title} {San Francisco:
  W.H.~Freeman and Co., 1973}}}\ (\bibinfo {year} {1973})\BibitemShut {NoStop}%
\bibitem [{\citenamefont {{Heggie}}\ and\ \citenamefont
  {{Rasio}}(1996)}]{Heggie_96}%
  \BibitemOpen
  \bibfield  {author} {\bibinfo {author} {\bibfnamefont {D.~C.}\ \bibnamefont
  {{Heggie}}}\ and\ \bibinfo {author} {\bibfnamefont {F.~A.}\ \bibnamefont
  {{Rasio}}},\ }\href {\doibase 10.1093/mnras/282.3.1064} {\bibfield  {journal}
  {\bibinfo  {journal} {\mnras}\ }\textbf {\bibinfo {volume} {282}},\ \bibinfo
  {pages} {1064} (\bibinfo {year} {1996})},\ \Eprint
  {http://arxiv.org/abs/astro-ph/9506082} {astro-ph/9506082} \BibitemShut
  {NoStop}%
\bibitem [{\citenamefont {{Quinlan}}(1996)}]{Quinlan_96}%
  \BibitemOpen
  \bibfield  {author} {\bibinfo {author} {\bibfnamefont {G.~D.}\ \bibnamefont
  {{Quinlan}}},\ }\href {\doibase 10.1016/S1384-1076(96)00003-6} {\bibfield
  {journal} {\bibinfo  {journal} {New Astron.}\ }\textbf {\bibinfo {volume}
  {1}},\ \bibinfo {pages} {35} (\bibinfo {year} {1996})},\ \Eprint
  {http://arxiv.org/abs/astro-ph/9601092} {astro-ph/9601092} \BibitemShut
  {NoStop}%
\bibitem [{\citenamefont {{Mack}}\ \emph {et~al.}(2007)\citenamefont {{Mack}},
  \citenamefont {{Ostriker}},\ and\ \citenamefont {{Ricotti}}}]{Mack_07}%
  \BibitemOpen
  \bibfield  {author} {\bibinfo {author} {\bibfnamefont {K.~J.}\ \bibnamefont
  {{Mack}}}, \bibinfo {author} {\bibfnamefont {J.~P.}\ \bibnamefont
  {{Ostriker}}}, \ and\ \bibinfo {author} {\bibfnamefont {M.}~\bibnamefont
  {{Ricotti}}},\ }\href {\doibase 10.1086/518998} {\bibfield  {journal}
  {\bibinfo  {journal} {\apj}\ }\textbf {\bibinfo {volume} {665}},\ \bibinfo
  {pages} {1277} (\bibinfo {year} {2007})},\ \Eprint
  {http://arxiv.org/abs/astro-ph/0608642} {astro-ph/0608642} \BibitemShut
  {NoStop}%
\bibitem [{\citenamefont {{Ricotti}}(2007)}]{Ricotti_07}%
  \BibitemOpen
  \bibfield  {author} {\bibinfo {author} {\bibfnamefont {M.}~\bibnamefont
  {{Ricotti}}},\ }\href {\doibase 10.1086/516562} {\bibfield  {journal}
  {\bibinfo  {journal} {\apj}\ }\textbf {\bibinfo {volume} {662}},\ \bibinfo
  {pages} {53} (\bibinfo {year} {2007})},\ \Eprint
  {http://arxiv.org/abs/0706.0864} {arXiv:0706.0864} \BibitemShut {NoStop}%
\bibitem [{\citenamefont {{Eroshenko}}(2016{\natexlab{b}})}]{Eroshenko_16b}%
  \BibitemOpen
  \bibfield  {author} {\bibinfo {author} {\bibfnamefont {Y.~N.}\ \bibnamefont
  {{Eroshenko}}},\ }\href {\doibase 10.1134/S1063773716060013} {\bibfield
  {journal} {\bibinfo  {journal} {Astronomy Letters}\ }\textbf {\bibinfo
  {volume} {42}},\ \bibinfo {pages} {347} (\bibinfo {year}
  {2016}{\natexlab{b}})},\ \Eprint {http://arxiv.org/abs/1607.00612}
  {arXiv:1607.00612 [astro-ph.HE]} \BibitemShut {NoStop}%
\bibitem [{\citenamefont {{Lacki}}\ and\ \citenamefont
  {{Beacom}}(2010)}]{Lacki_10}%
  \BibitemOpen
  \bibfield  {author} {\bibinfo {author} {\bibfnamefont {B.~C.}\ \bibnamefont
  {{Lacki}}}\ and\ \bibinfo {author} {\bibfnamefont {J.~F.}\ \bibnamefont
  {{Beacom}}},\ }\href {\doibase 10.1088/2041-8205/720/1/L67} {\bibfield
  {journal} {\bibinfo  {journal} {\apjl}\ }\textbf {\bibinfo {volume} {720}},\
  \bibinfo {pages} {L67} (\bibinfo {year} {2010})},\ \Eprint
  {http://arxiv.org/abs/1003.3466} {arXiv:1003.3466} \BibitemShut {NoStop}%
\bibitem [{\citenamefont {{Hayasaki}}(2009)}]{Hayasaki_09}%
  \BibitemOpen
  \bibfield  {author} {\bibinfo {author} {\bibfnamefont {K.}~\bibnamefont
  {{Hayasaki}}},\ }\href {\doibase 10.1093/pasj/61.1.65} {\bibfield  {journal}
  {\bibinfo  {journal} {\pasj}\ }\textbf {\bibinfo {volume} {61}},\ \bibinfo
  {pages} {65} (\bibinfo {year} {2009})},\ \Eprint
  {http://arxiv.org/abs/0805.3408} {arXiv:0805.3408} \BibitemShut {NoStop}%
\bibitem [{\citenamefont {{Abramowicz}}\ and\ \citenamefont
  {{Fragile}}(2013)}]{Abramowicz_13}%
  \BibitemOpen
  \bibfield  {author} {\bibinfo {author} {\bibfnamefont {M.~A.}\ \bibnamefont
  {{Abramowicz}}}\ and\ \bibinfo {author} {\bibfnamefont {P.~C.}\ \bibnamefont
  {{Fragile}}},\ }\href {\doibase 10.12942/lrr-2013-1} {\bibfield  {journal}
  {\bibinfo  {journal} {Living Reviews in Relativity}\ }\textbf {\bibinfo
  {volume} {16}},\ \bibinfo {eid} {1} (\bibinfo {year} {2013})},\ \Eprint
  {http://arxiv.org/abs/1104.5499} {arXiv:1104.5499 [astro-ph.HE]} \BibitemShut
  {NoStop}%
\bibitem [{\citenamefont {{Hoyle}}\ and\ \citenamefont
  {{Lyttleton}}(1939)}]{Hoyle_39}%
  \BibitemOpen
  \bibfield  {author} {\bibinfo {author} {\bibfnamefont {F.}~\bibnamefont
  {{Hoyle}}}\ and\ \bibinfo {author} {\bibfnamefont {R.~A.}\ \bibnamefont
  {{Lyttleton}}},\ }\href {\doibase 10.1017/S0305004100021150} {\bibfield
  {journal} {\bibinfo  {journal} {Proceedings of the Cambridge Philosophical
  Society}\ }\textbf {\bibinfo {volume} {35}},\ \bibinfo {pages} {405}
  (\bibinfo {year} {1939})}\BibitemShut {NoStop}%
\bibitem [{\citenamefont {{Bondi}}\ and\ \citenamefont
  {{Hoyle}}(1944)}]{Bondi_44}%
  \BibitemOpen
  \bibfield  {author} {\bibinfo {author} {\bibfnamefont {H.}~\bibnamefont
  {{Bondi}}}\ and\ \bibinfo {author} {\bibfnamefont {F.}~\bibnamefont
  {{Hoyle}}},\ }\href {\doibase 10.1093/mnras/104.5.273} {\bibfield  {journal}
  {\bibinfo  {journal} {\mnras}\ }\textbf {\bibinfo {volume} {104}},\ \bibinfo
  {pages} {273} (\bibinfo {year} {1944})}\BibitemShut {NoStop}%
\bibitem [{\citenamefont {{Bondi}}(1952)}]{Bondi_52}%
  \BibitemOpen
  \bibfield  {author} {\bibinfo {author} {\bibfnamefont {H.}~\bibnamefont
  {{Bondi}}},\ }\href {\doibase 10.1093/mnras/112.2.195} {\bibfield  {journal}
  {\bibinfo  {journal} {\mnras}\ }\textbf {\bibinfo {volume} {112}},\ \bibinfo
  {pages} {195} (\bibinfo {year} {1952})}\BibitemShut {NoStop}%
\bibitem [{\citenamefont {{Artymowicz}}\ and\ \citenamefont
  {{Lubow}}(1994)}]{Artymowicz_94}%
  \BibitemOpen
  \bibfield  {author} {\bibinfo {author} {\bibfnamefont {P.}~\bibnamefont
  {{Artymowicz}}}\ and\ \bibinfo {author} {\bibfnamefont {S.~H.}\ \bibnamefont
  {{Lubow}}},\ }\href {\doibase 10.1086/173679} {\bibfield  {journal} {\bibinfo
   {journal} {\apj}\ }\textbf {\bibinfo {volume} {421}},\ \bibinfo {pages}
  {651} (\bibinfo {year} {1994})}\BibitemShut {NoStop}%
\bibitem [{\citenamefont {{Goldreich}}\ and\ \citenamefont
  {{Tremaine}}(1979)}]{Goldreich_79}%
  \BibitemOpen
  \bibfield  {author} {\bibinfo {author} {\bibfnamefont {P.}~\bibnamefont
  {{Goldreich}}}\ and\ \bibinfo {author} {\bibfnamefont {S.}~\bibnamefont
  {{Tremaine}}},\ }\href {\doibase 10.1086/157448} {\bibfield  {journal}
  {\bibinfo  {journal} {\apj}\ }\textbf {\bibinfo {volume} {233}},\ \bibinfo
  {pages} {857} (\bibinfo {year} {1979})}\BibitemShut {NoStop}%
\bibitem [{\citenamefont {{Shakura}}\ and\ \citenamefont
  {{Sunyaev}}(1973)}]{Shakura_73}%
  \BibitemOpen
  \bibfield  {author} {\bibinfo {author} {\bibfnamefont {N.~I.}\ \bibnamefont
  {{Shakura}}}\ and\ \bibinfo {author} {\bibfnamefont {R.~A.}\ \bibnamefont
  {{Sunyaev}}},\ }\href@noop {} {\bibfield  {journal} {\bibinfo  {journal}
  {\aap}\ }\textbf {\bibinfo {volume} {24}},\ \bibinfo {pages} {337} (\bibinfo
  {year} {1973})}\BibitemShut {NoStop}%
\bibitem [{\citenamefont {{Pringle}}(1981)}]{Pringle_81}%
  \BibitemOpen
  \bibfield  {author} {\bibinfo {author} {\bibfnamefont {J.~E.}\ \bibnamefont
  {{Pringle}}},\ }\href {\doibase 10.1146/annurev.aa.19.090181.001033}
  {\bibfield  {journal} {\bibinfo  {journal} {\araa}\ }\textbf {\bibinfo
  {volume} {19}},\ \bibinfo {pages} {137} (\bibinfo {year} {1981})}\BibitemShut
  {NoStop}%
\bibitem [{\citenamefont {{Hayasaki}}\ \emph {et~al.}(2013)\citenamefont
  {{Hayasaki}}, \citenamefont {{Yagi}}, \citenamefont {{Tanaka}},\ and\
  \citenamefont {{Mineshige}}}]{Hayasaki_13}%
  \BibitemOpen
  \bibfield  {author} {\bibinfo {author} {\bibfnamefont {K.}~\bibnamefont
  {{Hayasaki}}}, \bibinfo {author} {\bibfnamefont {K.}~\bibnamefont {{Yagi}}},
  \bibinfo {author} {\bibfnamefont {T.}~\bibnamefont {{Tanaka}}}, \ and\
  \bibinfo {author} {\bibfnamefont {S.}~\bibnamefont {{Mineshige}}},\ }\href
  {\doibase 10.1103/PhysRevD.87.044051} {\bibfield  {journal} {\bibinfo
  {journal} {\prd}\ }\textbf {\bibinfo {volume} {87}},\ \bibinfo {eid} {044051}
  (\bibinfo {year} {2013})},\ \Eprint {http://arxiv.org/abs/1201.2858}
  {arXiv:1201.2858 [astro-ph.CO]} \BibitemShut {NoStop}%
\bibitem [{\citenamefont {{MacFadyen}}\ and\ \citenamefont
  {{Milosavljevi{\'c}}}(2008)}]{MacFadyen_08}%
  \BibitemOpen
  \bibfield  {author} {\bibinfo {author} {\bibfnamefont {A.~I.}\ \bibnamefont
  {{MacFadyen}}}\ and\ \bibinfo {author} {\bibfnamefont {M.}~\bibnamefont
  {{Milosavljevi{\'c}}}},\ }\href {\doibase 10.1086/523869} {\bibfield
  {journal} {\bibinfo  {journal} {\apj}\ }\textbf {\bibinfo {volume} {672}},\
  \bibinfo {eid} {83-93} (\bibinfo {year} {2008})},\ \Eprint
  {http://arxiv.org/abs/astro-ph/0607467} {astro-ph/0607467} \BibitemShut
  {NoStop}%
\bibitem [{\citenamefont {{Tang}}\ \emph {et~al.}(2017)\citenamefont {{Tang}},
  \citenamefont {{MacFadyen}},\ and\ \citenamefont {{Haiman}}}]{Tang_17}%
  \BibitemOpen
  \bibfield  {author} {\bibinfo {author} {\bibfnamefont {Y.}~\bibnamefont
  {{Tang}}}, \bibinfo {author} {\bibfnamefont {A.}~\bibnamefont {{MacFadyen}}},
  \ and\ \bibinfo {author} {\bibfnamefont {Z.}~\bibnamefont {{Haiman}}},\
  }\href {\doibase 10.1093/mnras/stx1130} {\bibfield  {journal} {\bibinfo
  {journal} {\mnras}\ }\textbf {\bibinfo {volume} {469}},\ \bibinfo {pages}
  {4258} (\bibinfo {year} {2017})},\ \Eprint {http://arxiv.org/abs/1703.03913}
  {arXiv:1703.03913 [astro-ph.HE]} \BibitemShut {NoStop}%
\bibitem [{\citenamefont {{Quinlan}}\ and\ \citenamefont
  {{Shapiro}}(1989)}]{Quinlan_89}%
  \BibitemOpen
  \bibfield  {author} {\bibinfo {author} {\bibfnamefont {G.~D.}\ \bibnamefont
  {{Quinlan}}}\ and\ \bibinfo {author} {\bibfnamefont {S.~L.}\ \bibnamefont
  {{Shapiro}}},\ }\href {\doibase 10.1086/167745} {\bibfield  {journal}
  {\bibinfo  {journal} {\apj}\ }\textbf {\bibinfo {volume} {343}},\ \bibinfo
  {pages} {725} (\bibinfo {year} {1989})}\BibitemShut {NoStop}%
\bibitem [{\citenamefont {{Prada}}\ \emph {et~al.}(2012)\citenamefont
  {{Prada}}, \citenamefont {{Klypin}}, \citenamefont {{Cuesta}}, \citenamefont
  {{Betancort-Rijo}},\ and\ \citenamefont {{Primack}}}]{Prada_12}%
  \BibitemOpen
  \bibfield  {author} {\bibinfo {author} {\bibfnamefont {F.}~\bibnamefont
  {{Prada}}}, \bibinfo {author} {\bibfnamefont {A.~A.}\ \bibnamefont
  {{Klypin}}}, \bibinfo {author} {\bibfnamefont {A.~J.}\ \bibnamefont
  {{Cuesta}}}, \bibinfo {author} {\bibfnamefont {J.~E.}\ \bibnamefont
  {{Betancort-Rijo}}}, \ and\ \bibinfo {author} {\bibfnamefont
  {J.}~\bibnamefont {{Primack}}},\ }\href {\doibase
  10.1111/j.1365-2966.2012.21007.x} {\bibfield  {journal} {\bibinfo  {journal}
  {\mnras}\ }\textbf {\bibinfo {volume} {423}},\ \bibinfo {pages} {3018}
  (\bibinfo {year} {2012})},\ \Eprint {http://arxiv.org/abs/1104.5130}
  {arXiv:1104.5130} \BibitemShut {NoStop}%
\bibitem [{\citenamefont {{Tinker}}\ \emph {et~al.}(2008)\citenamefont
  {{Tinker}}, \citenamefont {{Kravtsov}}, \citenamefont {{Klypin}},
  \citenamefont {{Abazajian}}, \citenamefont {{Warren}}, \citenamefont
  {{Yepes}}, \citenamefont {{Gottl{\"o}ber}},\ and\ \citenamefont
  {{Holz}}}]{Tinker_08}%
  \BibitemOpen
  \bibfield  {author} {\bibinfo {author} {\bibfnamefont {J.}~\bibnamefont
  {{Tinker}}}, \bibinfo {author} {\bibfnamefont {A.~V.}\ \bibnamefont
  {{Kravtsov}}}, \bibinfo {author} {\bibfnamefont {A.}~\bibnamefont
  {{Klypin}}}, \bibinfo {author} {\bibfnamefont {K.}~\bibnamefont
  {{Abazajian}}}, \bibinfo {author} {\bibfnamefont {M.}~\bibnamefont
  {{Warren}}}, \bibinfo {author} {\bibfnamefont {G.}~\bibnamefont {{Yepes}}},
  \bibinfo {author} {\bibfnamefont {S.}~\bibnamefont {{Gottl{\"o}ber}}}, \ and\
  \bibinfo {author} {\bibfnamefont {D.~E.}\ \bibnamefont {{Holz}}},\ }\href
  {\doibase 10.1086/591439} {\bibfield  {journal} {\bibinfo  {journal} {\apj}\
  }\textbf {\bibinfo {volume} {688}},\ \bibinfo {eid} {709-728} (\bibinfo
  {year} {2008})},\ \Eprint {http://arxiv.org/abs/0803.2706} {arXiv:0803.2706}
  \BibitemShut {NoStop}%
\bibitem [{\citenamefont {{Press}}\ and\ \citenamefont
  {{Schechter}}(1974)}]{PS}%
  \BibitemOpen
  \bibfield  {author} {\bibinfo {author} {\bibfnamefont {W.~H.}\ \bibnamefont
  {{Press}}}\ and\ \bibinfo {author} {\bibfnamefont {P.}~\bibnamefont
  {{Schechter}}},\ }\href {\doibase 10.1086/152650} {\bibfield  {journal}
  {\bibinfo  {journal} {\apj}\ }\textbf {\bibinfo {volume} {187}},\ \bibinfo
  {pages} {425} (\bibinfo {year} {1974})}\BibitemShut {NoStop}%
\bibitem [{\citenamefont {{Binney}}\ and\ \citenamefont
  {{Tremaine}}(2008)}]{BT}%
  \BibitemOpen
  \bibfield  {author} {\bibinfo {author} {\bibfnamefont {J.}~\bibnamefont
  {{Binney}}}\ and\ \bibinfo {author} {\bibfnamefont {S.}~\bibnamefont
  {{Tremaine}}},\ }\href@noop {} {\emph {\bibinfo {title} {Galactic Dynamics:
  Second Edition}}}\ (\bibinfo  {publisher} {Princeton University Press},\
  \bibinfo {year} {2008})\BibitemShut {NoStop}%
\bibitem [{\citenamefont {{Afshordi}}\ \emph {et~al.}(2003)\citenamefont
  {{Afshordi}}, \citenamefont {{McDonald}},\ and\ \citenamefont
  {{Spergel}}}]{Afshordi_03}%
  \BibitemOpen
  \bibfield  {author} {\bibinfo {author} {\bibfnamefont {N.}~\bibnamefont
  {{Afshordi}}}, \bibinfo {author} {\bibfnamefont {P.}~\bibnamefont
  {{McDonald}}}, \ and\ \bibinfo {author} {\bibfnamefont {D.~N.}\ \bibnamefont
  {{Spergel}}},\ }\href {\doibase 10.1086/378763} {\bibfield  {journal}
  {\bibinfo  {journal} {\apjl}\ }\textbf {\bibinfo {volume} {594}},\ \bibinfo
  {pages} {L71} (\bibinfo {year} {2003})},\ \Eprint
  {http://arxiv.org/abs/astro-ph/0302035} {astro-ph/0302035} \BibitemShut
  {NoStop}%
\bibitem [{\citenamefont {Nishikawa}\ \emph {et~al.}(2017)\citenamefont
  {Nishikawa}, \citenamefont {Kovetz}, \citenamefont {Kamionkowski},\ and\
  \citenamefont {Silk}}]{Nishikawa:2017chy}%
  \BibitemOpen
  \bibfield  {author} {\bibinfo {author} {\bibfnamefont {H.}~\bibnamefont
  {Nishikawa}}, \bibinfo {author} {\bibfnamefont {E.~D.}\ \bibnamefont
  {Kovetz}}, \bibinfo {author} {\bibfnamefont {M.}~\bibnamefont
  {Kamionkowski}}, \ and\ \bibinfo {author} {\bibfnamefont {J.}~\bibnamefont
  {Silk}},\ }\href@noop {} {\  (\bibinfo {year} {2017})},\ \Eprint
  {http://arxiv.org/abs/1708.08449} {arXiv:1708.08449 [astro-ph.CO]}
  \BibitemShut {NoStop}%
\bibitem [{\citenamefont {Abbott}\ \emph
  {et~al.}(2017{\natexlab{b}})\citenamefont {Abbott} \emph {et~al.}}]{LIGO_17}%
  \BibitemOpen
  \bibfield  {author} {\bibinfo {author} {\bibfnamefont {B.~P.}\ \bibnamefont
  {Abbott}} \emph {et~al.} (\bibinfo {collaboration} {LIGO Scientific
  Collaboration and Virgo Collaboration}),\ }\href@noop {} {\bibfield
  {journal} {\bibinfo  {journal} {ArXiv e-prints}\ } (\bibinfo {year}
  {2017}{\natexlab{b}})},\ \Eprint {http://arxiv.org/abs/1704.04628}
  {arXiv:1704.04628 [gr-qc]} \BibitemShut {NoStop}%
\bibitem [{\citenamefont {Abbott}\ \emph
  {et~al.}(2016{\natexlab{c}})\citenamefont {Abbott} \emph
  {et~al.}}]{LIGO_16bS}%
  \BibitemOpen
  \bibfield  {author} {\bibinfo {author} {\bibfnamefont {B.~P.}\ \bibnamefont
  {Abbott}} \emph {et~al.} (\bibinfo {collaboration} {LIGO Scientific
  Collaboration and Virgo Collaboration}),\ }\href {\doibase
  10.3847/0067-0049/227/2/14} {\bibfield  {journal} {\bibinfo  {journal}
  {\apjs}\ }\textbf {\bibinfo {volume} {227}},\ \bibinfo {eid} {14} (\bibinfo
  {year} {2016}{\natexlab{c}})},\ \Eprint {http://arxiv.org/abs/1606.03939}
  {arXiv:1606.03939 [astro-ph.HE]} \BibitemShut {NoStop}%
\bibitem [{\citenamefont {Abbott}\ \emph
  {et~al.}(2016{\natexlab{d}})\citenamefont {Abbott} \emph
  {et~al.}}]{LIGO_16b}%
  \BibitemOpen
  \bibfield  {author} {\bibinfo {author} {\bibfnamefont {B.~P.}\ \bibnamefont
  {Abbott}} \emph {et~al.} (\bibinfo {collaboration} {LIGO Scientific
  Collaboration and Virgo Collaboration}),\ }\href {\doibase
  10.3847/2041-8205/833/1/L1} {\bibfield  {journal} {\bibinfo  {journal}
  {\apjl}\ }\textbf {\bibinfo {volume} {833}},\ \bibinfo {eid} {L1} (\bibinfo
  {year} {2016}{\natexlab{d}})},\ \Eprint {http://arxiv.org/abs/1602.03842}
  {arXiv:1602.03842 [astro-ph.HE]} \BibitemShut {NoStop}%
\bibitem [{\citenamefont {{Hawkins}}(2015)}]{Hawkins_15}%
  \BibitemOpen
  \bibfield  {author} {\bibinfo {author} {\bibfnamefont {M.~R.~S.}\
  \bibnamefont {{Hawkins}}},\ }\href {\doibase 10.1051/0004-6361/201425400}
  {\bibfield  {journal} {\bibinfo  {journal} {\aap}\ }\textbf {\bibinfo
  {volume} {575}},\ \bibinfo {eid} {A107} (\bibinfo {year} {2015})},\ \Eprint
  {http://arxiv.org/abs/1503.01935} {arXiv:1503.01935} \BibitemShut {NoStop}%
\bibitem [{\citenamefont {{Chandrasekhar}}(1943)}]{Chandra_43}%
  \BibitemOpen
  \bibfield  {author} {\bibinfo {author} {\bibfnamefont {S.}~\bibnamefont
  {{Chandrasekhar}}},\ }\href {\doibase 10.1103/RevModPhys.15.1} {\bibfield
  {journal} {\bibinfo  {journal} {Reviews of Modern Physics}\ }\textbf
  {\bibinfo {volume} {15}},\ \bibinfo {pages} {1} (\bibinfo {year}
  {1943})}\BibitemShut {NoStop}%
\end{thebibliography}%
 
\end{document}